\documentclass[postprint,superscriptaddress,showpacs,preprintnumbers,nobibnotes,amsmath,amssymb,aps,longbibliography]{revtex4-2}

\usepackage[T1]{fontenc}
\usepackage[colorlinks=true,citecolor=blue,linkcolor=blue]{hyperref}
\usepackage{cleveref}
\usepackage{booktabs}
\usepackage{graphicx}
\usepackage{siunitx}
\usepackage{enumitem}
\usepackage{multirow}
\usepackage{pifont}
\usepackage[dvipsnames]{xcolor}
\usepackage[normalem]{ulem}
\usepackage{xspace}
\usepackage{geometry}
\usepackage{setspace}

\newif\ifdraft

\ifdraft
  \usepackage{lineno}
\fi

\begin{document}

\ifdraft
  \linenumbers
\fi

\onehalfspacing

\title{High-Energy Neutrino Tomography of the Earth's Interior with IceCube}

\affiliation{III. Physikalisches Institut, RWTH Aachen University, D-52056 Aachen, Germany}
\affiliation{Department of Physics, University of Adelaide, Adelaide, 5005, Australia}
\affiliation{Dept. of Physics and Astronomy, University of Alaska Anchorage, 3211 Providence Dr., Anchorage, AK 99508, USA}
\affiliation{School of Physics and Center for Relativistic Astrophysics, Georgia Institute of Technology, Atlanta, GA 30332, USA}
\affiliation{Dept. of Physics, Southern University, Baton Rouge, LA 70813, USA}
\affiliation{Dept. of Physics, University of California, Berkeley, CA 94720, USA}
\affiliation{Lawrence Berkeley National Laboratory, Berkeley, CA 94720, USA}
\affiliation{Institut f{\"u}r Physik, Humboldt-Universit{\"a}t zu Berlin, D-12489 Berlin, Germany}
\affiliation{Fakult{\"a}t f{\"u}r Physik {\&} Astronomie, Ruhr-Universit{\"a}t Bochum, D-44780 Bochum, Germany}
\affiliation{Universit{\'e} Libre de Bruxelles, Science Faculty CP230, B-1050 Brussels, Belgium}
\affiliation{Vrije Universiteit Brussel (VUB), Dienst ELEM, B-1050 Brussels, Belgium}
\affiliation{Dept. of Physics, Simon Fraser University, Burnaby, BC V5A 1S6, Canada}
\affiliation{Department of Physics and Laboratory for Particle Physics and Cosmology, Harvard University, Cambridge, MA 02138, USA}
\affiliation{Dept. of Physics, Massachusetts Institute of Technology, Cambridge, MA 02139, USA}
\affiliation{Dept. of Physics and The International Center for Hadron Astrophysics, Chiba University, Chiba 263-8522, Japan}
\affiliation{Department of Physics, Loyola University Chicago, Chicago, IL 60660, USA}
\affiliation{Dept. of Physics and Astronomy, University of Canterbury, Private Bag 4800, Christchurch, New Zealand}
\affiliation{Dept. of Physics, University of Maryland, College Park, MD 20742, USA}
\affiliation{Dept. of Astronomy, Ohio State University, Columbus, OH 43210, USA}
\affiliation{Dept. of Physics and Center for Cosmology and Astro-Particle Physics, Ohio State University, Columbus, OH 43210, USA}
\affiliation{Niels Bohr Institute, University of Copenhagen, DK-2100 Copenhagen, Denmark}
\affiliation{Dept. of Physics, TU Dortmund University, D-44221 Dortmund, Germany}
\affiliation{Dept. of Physics and Astronomy, Michigan State University, East Lansing, MI 48824, USA}
\affiliation{Dept. of Physics, University of Alberta, Edmonton, Alberta, T6G 2E1, Canada}
\affiliation{Erlangen Centre for Astroparticle Physics, Friedrich-Alexander-Universit{\"a}t Erlangen-N{\"u}rnberg, D-91058 Erlangen, Germany}
\affiliation{Physik-department, Technische Universit{\"a}t M{\"u}nchen, D-85748 Garching, Germany}
\affiliation{D{\'e}partement de physique nucl{\'e}aire et corpusculaire, Universit{\'e} de Gen{\`e}ve, CH-1211 Gen{\`e}ve, Switzerland}
\affiliation{Dept. of Physics and Astronomy, University of Gent, B-9000 Gent, Belgium}
\affiliation{Dept. of Physics and Astronomy, University of California, Irvine, CA 92697, USA}
\affiliation{Karlsruhe Institute of Technology, Institute for Astroparticle Physics, D-76021 Karlsruhe, Germany}
\affiliation{Karlsruhe Institute of Technology, Institute of Experimental Particle Physics, D-76021 Karlsruhe, Germany}
\affiliation{Dept. of Physics, Engineering Physics, and Astronomy, Queen's University, Kingston, ON K7L 3N6, Canada}
\affiliation{Department of Physics {\&} Astronomy, University of Nevada, Las Vegas, NV 89154, USA}
\affiliation{Nevada Center for Astrophysics, University of Nevada, Las Vegas, NV 89154, USA}
\affiliation{Dept. of Physics and Astronomy, University of Kansas, Lawrence, KS 66045, USA}
\affiliation{UCLouvain, Centre for Cosmology, Particle Physics and Phenomenology, CP3, Chemin du Cyclotron 2, 1348 Louvain-la-Neuve, Belgium}
\affiliation{Department of Physics, Mercer University, Macon, GA 31207-0001, USA}
\affiliation{Dept. of Astronomy, University of Wisconsin{\textemdash}Madison, Madison, WI 53706, USA}
\affiliation{Dept. of Physics and Wisconsin IceCube Particle Astrophysics Center, University of Wisconsin{\textemdash}Madison, Madison, WI 53706, USA}
\affiliation{Institute of Physics, University of Mainz, Staudinger Weg 7, D-55099 Mainz, Germany}
\affiliation{Department of Physics, Marquette University, Milwaukee, WI 53201, USA}
\affiliation{Institut f{\"u}r Kernphysik, Universit{\"a}t M{\"u}nster, D-48149 M{\"u}nster, Germany}
\affiliation{Bartol Research Institute and Dept. of Physics and Astronomy, University of Delaware, Newark, DE 19716, USA}
\affiliation{Dept. of Physics, Yale University, New Haven, CT 06520, USA}
\affiliation{Columbia Astrophysics and Nevis Laboratories, Columbia University, New York, NY 10027, USA}
\affiliation{Dept. of Physics, University of Oxford, Parks Road, Oxford OX1 3PU, United Kingdom}
\affiliation{Dipartimento di Fisica e Astronomia Galileo Galilei, Universit{\`a} Degli Studi di Padova, I-35122 Padova PD, Italy}
\affiliation{Dept. of Physics, Drexel University, 3141 Chestnut Street, Philadelphia, PA 19104, USA}
\affiliation{Physics Department, South Dakota School of Mines and Technology, Rapid City, SD 57701, USA}
\affiliation{Dept. of Physics, University of Wisconsin, River Falls, WI 54022, USA}
\affiliation{Dept. of Physics and Astronomy, University of Rochester, Rochester, NY 14627, USA}
\affiliation{Department of Physics and Astronomy, University of Utah, Salt Lake City, UT 84112, USA}
\affiliation{Dept. of Physics, Chung-Ang University, Seoul 06974, Republic of Korea}
\affiliation{Oskar Klein Centre and Dept. of Physics, Stockholm University, SE-10691 Stockholm, Sweden}
\affiliation{Dept. of Physics and Astronomy, Stony Brook University, Stony Brook, NY 11794-3800, USA}
\affiliation{Dept. of Physics, Sungkyunkwan University, Suwon 16419, Republic of Korea}
\affiliation{Institute of Physics, Academia Sinica, Taipei, 11529, Taiwan}
\affiliation{Dept. of Physics and Astronomy, University of Alabama, Tuscaloosa, AL 35487, USA}
\affiliation{Dept. of Astronomy and Astrophysics, Pennsylvania State University, University Park, PA 16802, USA}
\affiliation{Dept. of Physics, Pennsylvania State University, University Park, PA 16802, USA}
\affiliation{Dept. of Physics and Astronomy, Uppsala University, Box 516, SE-75120 Uppsala, Sweden}
\affiliation{Dept. of Physics, University of Wuppertal, D-42119 Wuppertal, Germany}
\affiliation{Deutsches Elektronen-Synchrotron DESY, Platanenallee 6, D-15738 Zeuthen, Germany}

\author{R. Abbasi}
\affiliation{Department of Physics, Loyola University Chicago, Chicago, IL 60660, USA}
\author{M. Ackermann}
\affiliation{Deutsches Elektronen-Synchrotron DESY, Platanenallee 6, D-15738 Zeuthen, Germany}
\author{J. Adams}
\affiliation{Dept. of Physics and Astronomy, University of Canterbury, Private Bag 4800, Christchurch, New Zealand}
\author{J. A. Aguilar}
\affiliation{Universit{\'e} Libre de Bruxelles, Science Faculty CP230, B-1050 Brussels, Belgium}
\author{M. Ahlers}
\affiliation{Niels Bohr Institute, University of Copenhagen, DK-2100 Copenhagen, Denmark}
\author{J.M. Alameddine}
\affiliation{Dept. of Physics, TU Dortmund University, D-44221 Dortmund, Germany}
\author{S. Ali}
\affiliation{Dept. of Physics and Astronomy, University of Kansas, Lawrence, KS 66045, USA}
\author{N. M. Amin}
\affiliation{Bartol Research Institute and Dept. of Physics and Astronomy, University of Delaware, Newark, DE 19716, USA}
\author{K. Andeen}
\affiliation{Department of Physics, Marquette University, Milwaukee, WI 53201, USA}
\author{C. Arg{\"u}elles}
\affiliation{Department of Physics and Laboratory for Particle Physics and Cosmology, Harvard University, Cambridge, MA 02138, USA}
% ...
\date{\today}

\collaboration{IceCube Collaboration}
\noaffiliation

\maketitle


\textbf{The Earth’s interior reflects its geological evolution, from accretion history to present-day dynamics. 
Its structure drives the geodynamo in the outer core, generating the magnetic field that shields surface life from charged cosmic radiation and may help explain why similar planets such as Mars lack such a field.
The Earth's central observables are the radial distribution of matter and derived quantities such as the Earth's mass and moment of inertia.
To date, these have been inferred primarily from gravity and seismic wave propagation, which probe the macroscopic response of matter to elastic and gravitational forces.
Here we instead probe the Earth’s density profile using high-energy neutrinos observed by the IceCube Neutrino Observatory located at the South Pole. 
We use 10.7 years of data consisting of predominantly muon-neutrino events in the energy range 500\,GeV--100\,TeV. 
This includes both the copious atmospheric neutrino flux, produced by cosmic-ray interactions in the Earth's atmosphere, and the much rarer diffuse astrophysical neutrino flux.
The attenuation of the neutrino flux depends on the traversed column density and the neutrino energy.
By measuring the zenith- and energy-dependent flux suppression we infer a radial Earth density profile by fitting a concentric uniform-density shell model that accounts for flux, cross section, detector efficiency and glacial-ice systematic uncertainties.
From the resulting density posteriors, we derive the Earth's mass and polar moment of inertia as \emph{measured} by neutrinos.
These measurements are the most precise to date based on the weak interaction and are consistent with the Preliminary Reference Earth Model and gravitational determinations. 
Our results demonstrate that neutrinos provide a novel probe of planetary interiors via a distinct physical interaction, complementing gravity and seismology. 
With improved detectors and precision, neutrinos will further contribute to a multifaceted understanding of the Earth's structure.
}

The most precise probes of the Earth's deep interior have long relied on seismic waves; inferring density using seismic velocities has been explored since the 1920s~\cite{williamson1923density,clark1964density}.
After nearly a century of study, a modern understanding of the Earth's density profile, using seismic information as well as normal modes of oscillation, has emerged as the Preliminary Reference Earth Model (PREM)~\cite{DZIEWONSKI1981297, Gilbert1975} and its contemporaries~\cite{AK135:1995}.
While broadly descriptive, these measurements ultimately infer structure from the macroscopic response of matter to gravity and elastic forces.
A probe via the weak force, and neutrinos, offers another possibility, sensitive to the Earth's density distribution through a fundamentally different interaction.

Neutrinos carry no electric charge and interact only via the weak force.
Their small interaction cross sections allow them to traverse Earth-scale matter distributions~\cite{Garcia:2020jwr,Reno:2023sdm}.
Because the neutrino--nucleon cross section increases with energy, the Earth becomes partially opaque for neutrinos above $\sim 10\,\mathrm{TeV}$; the resulting attenuation depends on the column density along the trajectory~\cite{Gonzalez-Garcia:2007wfs,Vincent:2017svp}.
This energy scale marks the transition from a near-transparent Earth, where most neutrinos traverse the planet unscathed, to a regime where interactions during transit become common enough to measurably suppress the flux of neutrinos.
In this regime, density changes affect not only the overall rate of neutrino events which traverse the Earth, but also the \emph{shape} of the observed distribution in energy and zenith angle (see Fig.~\ref{fig:schematic}): the most energetic neutrinos with the longest trajectories through the Earth (more vertical) are attenuated most strongly.

This shape information also provides some leverage to separate density changes in different radial layers from a total density normalization.
By measuring the energy- and zenith-dependent neutrino flux after propagation through the Earth and comparing it with expectations at production, we can infer the traversed column density and thereby constrain the radial density distribution using primarily the weak interaction.
Conversely, if the Earth's density profile is treated as known (e.g.\ from PREM), the same attenuation signal can instead be used to measure the neutrino--nucleon cross section~\cite{IceCube:2020rnc}.
The same density posterior yields derived global quantities such as the Earth's mass and moment of inertia as measured by neutrinos.

These measurements are enabled by neutrino telescopes such as the IceCube Neutrino Observatory~\cite{IceCube:2016zyt}, which can detect neutrinos above a few $\textrm{GeV}$ in energy.
Atmospheric neutrinos, a dominant source of neutrino signals at IceCube, provide a natural source for Earth tomography: they are produced when cosmic rays interact in the atmosphere, generating air showers with a large hadronic component that produces neutrinos, which propagate through the Earth.
Although the neutrino flux falls steeply with energy, approximately as $\sim E^{-3.7}$, it remains measurable at $\sim 1\,\textrm{TeV}$ and above, where attenuation becomes significant.
This is not a novel proposal: neutrino-attenuation tomography has been discussed for decades~\cite{Placci:1973yvk,Volkova:1974xa,Gonzalez-Garcia:2007wfs,Reynoso:2004dt,2012EGUGA}, but only recently have detector capabilities and event statistics enabled precise measurements.
The first data-driven determination of the Earth's density profile, mass, and moment of inertia using this approach was reported in Ref.~\cite{Donini:2018tsg}.
Here we present a next-generation measurement using an order of magnitude more data, and a significantly updated systematic uncertainty treatment at IceCube, to produce the most precise weak-interaction-based determination of the Earth's interior to date.

\section{Neutrino Physics at IceCube}\label{icecube}

\begin{figure}
    \centering
    \includegraphics[width=1.0\linewidth]{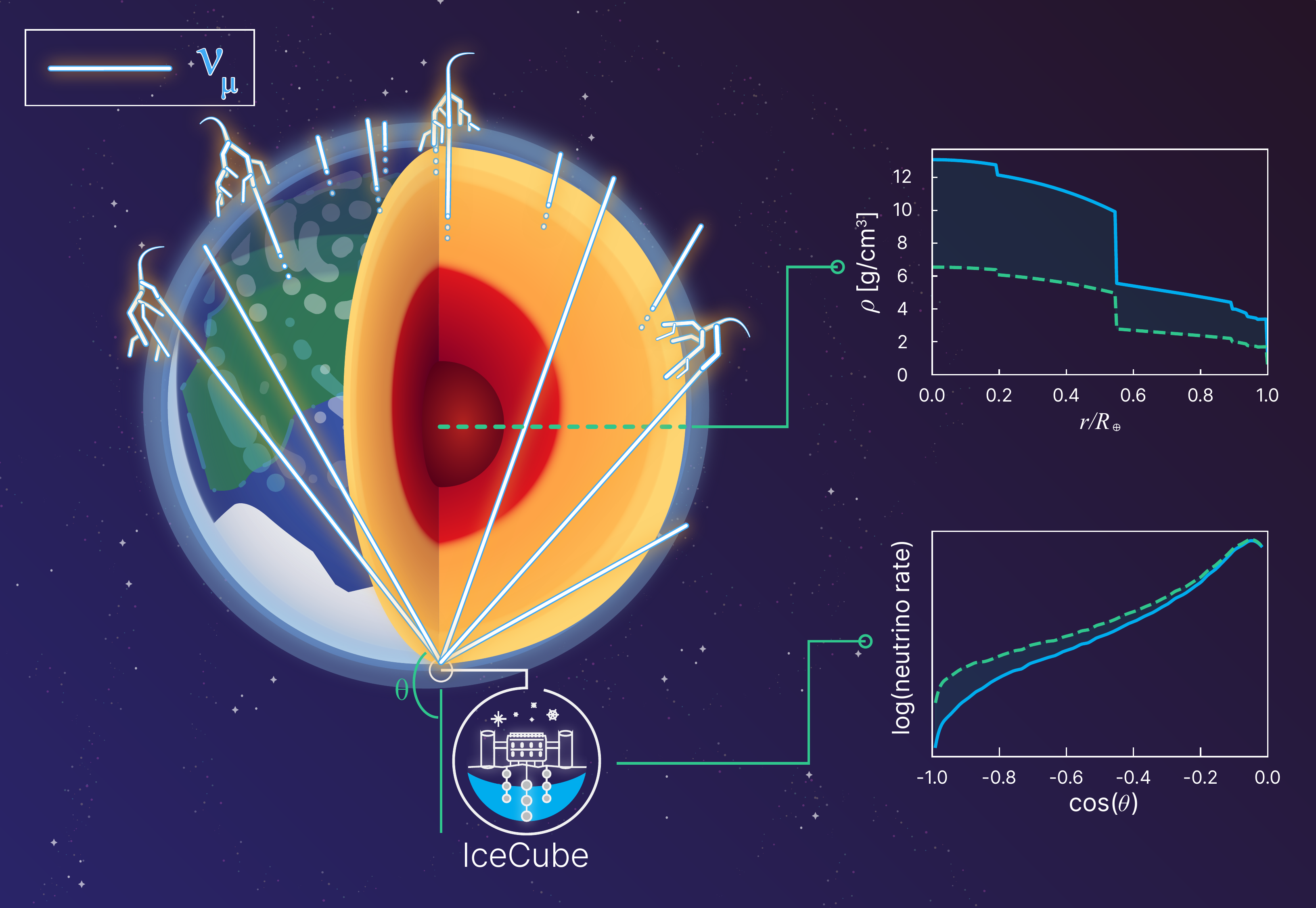}
    \caption{\textbf{A schematic of neutrinos being attenuated by the Earth, showing how density affects the neutrino rate.}
    The left part of the figure depicts an artist's rendition of the Earth with a cutaway view into the core, with neutrinos travelling towards IceCube at the South Pole represented as white lines.
    The zenith angle, $\theta$, used to describe the neutrino direction is also shown on the figure; it is the angle between the neutrino trajectory and the vertical at the South Pole.
    These neutrinos are produced in atmospheric cosmic ray showers, some of which are depicted.
    Some neutrinos are attenuated, represented by the dashed lines.
    In the top-right panel, we show the Earth's density, $\rho$, profile (solid line) as given by the PREM model, plotting density as a function of normalized radius $r/R_\oplus$, where $R_\oplus\approx 6371\,\mathrm{km}$. The dashed line shows a hypothetical alternate density profile at 50\% of the PREM density.
    The corresponding shift in the neutrino distribution (as a function of $\cos\theta$), exaggerated by a factor of 10 for illustrative purposes, is shown in the bottom-right panel; the solid and dashed lines correspond to the same two density profiles.
    The key quantitative takeaway is that as the density decreases, the neutrino rate increases, since fewer neutrinos are attenuated.
    By measuring the neutrino rate precisely, we can infer the density profile knowing the neutrino cross section in the Standard Model, sensitive not only to normalization effects but also to shape effects, since we probe different layers of the Earth by selecting different ranges of $\theta$.}
    \label{fig:schematic}
\end{figure}

IceCube is a neutrino detector that consists of a cubic-kilometer-sized volume of South Pole glacial ice instrumented with an array of $5{,}160$ photodetectors called digital optical modules (DOMs)~\cite{IceCube:2016zyt,IceCube:2010dpc}.
Each DOM consists of a 10-inch diameter photomultiplier tube along with supporting electronics housed in a spherical glass pressure vessel.
The role of this array of photodetectors is to detect the Cherenkov light induced by charged particles as they traverse the ice at speeds faster than the speed of light in that medium; these charged particles, which may come from neutrino interactions, give an important signature of the interaction, and allow us to reconstruct the energy and direction of the incoming neutrino.
In particular, at neutrino telescopes like IceCube, the direction is parametrized by the zenith angle, $\theta$, defined as the angle relative to the vertical at the South Pole. 
In Fig.~\ref{fig:schematic}, the zenith angle is labelled for one example neutrino trajectory.

Interpreting an energy- and zenith-dependent flux suppression in terms of Earth density requires, most importantly, a thorough understanding of the neutrino-nucleon cross sections that underpin each interaction.
The fundamental weak interaction is mediated by the $W^\pm$ and $Z^0$ gauge bosons which gain masses of ${\cal O}(10^2)$~GeV from spontaneous breaking of the governing $SU(2)_\text{L} \otimes U(1)_{Y}$ gauge symmetry by the Higgs mechanism, making neutrino-matter scattering a well-understood process in the Standard Model of particle physics~\cite{ParticleDataGroup:2024cfk}. 
Above energies of a few GeV, neutrinos ($\nu$) and antineutrinos ($\bar{\nu}$) interact with nucleons ($N$) mainly via \textit{deep inelastic scattering} (DIS) off a constituent parton (quark or gluon) by exchanging a $W^\pm$ (charged-current, CC) or a $Z^0$ (neutral-current, NC). The struck parton generates a hadronic shower and in the dominant CC process, the neutrino also converts to its corresponding charged lepton~\cite{ParticleDataGroup:2024cfk}. 
The differential cross-section is determined by the kinematic variables $Q^2$ (the relativistic invariant 4-momentum transfer), `Bjorken' $x$ (the longitudinal momentum fraction  of the nucleon carried by the struck parton), and $y$ (the `inelasticity' i.e. the fraction of the neutrino energy carried away by the hadronic shower)~\cite{Devenish:2004pb}:
\begin{equation}
\frac{d^2\sigma^{\text{CC,NC}}_{\nu,\bar{\nu}}}{d x~d y} = \frac{G^2_\text{F} M_N E_\nu}{\pi} \left(\frac{M_{W,Z}^2}{Q^2 + M_{W,Z}^2}\right)\left[\frac{1+(1-y)^2}{2}F_2^{\text{CC,NC}}(x, Q^2) - \frac{y^2}{2}F_L^{\text{CC,NC}}(x, Q^2) \pm y\left(1-\frac{y}{2}\right)xF_3^{\text{CC,NC}}(x, Q^2)\right].
\end{equation}
Here $G_\text{F}$ is Fermi's constant and $F_2$, $F_L$ and $F_3$ are the structure functions~\cite{Devenish:2004pb} determined by the parton distribution functions (PDF). 
In perturbative quantum chromodynamics (pQCD), these are given e.g. for $\nu$ at leading order by:
\begin{equation}
F_2 = x(u_\text{v} + d_\text{v} + 2s + 2b + \bar{u} + \bar{d} + 2 \bar{c}), \quad F_L = 0, \quad
xF_3 = x(u_\text{v} + d_\text{v} + 2s + 2b - \bar{u} - \bar{d} - 2 \bar{c}),
\end{equation}
where the subscript $\text{v}$ refers to the valence quarks (e.g. 2 $u$ quarks and 1 $d$ quark in the proton, 1 $u$ quark and 2 $d$ quarks in the neutron); while these dominate at high values of $x \lesssim 1$, as $x$ decreases the nucleon is increasingly dominated by \textit{sea} (i.e. virtual) quarks and gluons~\cite{ParticleDataGroup:2024cfk}. 
This evolution of the PDFs is described by the Dokshitzer-Gribov-Lipatov-Altarelli-Parisi (DGLAP) equations of pQCD~\cite{Dokshitzer:1977sg,Gribov:1972ri,Gribov:1972rt,Altarelli:1977zs}, taking as their input the PDFs which are extracted typically from analysis of electron-proton scattering, as well as other high energy experiments.

In the dominant CC process, the neutrino converts to its corresponding charged lepton with an accompanying hadronic shower.
From the propagator factor in Eq.~(1), most of the contribution to the cross section comes when $Q^2 \sim m_W^2$ and $x \sim m_W^2/(M_N E_\nu)$.
Hence as the neutrino energy increases, smaller values of $x$ are being probed and here the nucleons are dominated by sea quarks and gluons. 
Up to about a PeV, the relevant kinematic range in ($Q^2, x$) has been directly probed by the electron-proton scattering experiments performed at the HERA collider~\cite{H1:2009pze,H1:2015ubc}, while for higher neutrino energies, the PDFs need to be evolved using the DGLAP equations. 
Quantum effects are largely accounted for at next-to-leading order (NLO) i.e. 1-loop corrections, which must be calculated numerically. 
Even so, in the energy range probed here, the cross section can be determined to an accuracy of a few percent, taking into account the scatter between PDF determinations by different groups~\citep{Cooper-Sarkar:2011jtt}.
This well-constrained cross section is what enables the observed neutrino attenuation to be interpreted as a measurement of the traversed matter density.

With these specific interactions in mind, we now turn to the population of neutrinos that produce the observable signal.
Atmospheric neutrino production begins with cosmic rays.
They collide with nuclei in the Earth's atmosphere, producing air showers with hadronic components that generate mostly muon neutrinos for the energies considered in this work.
These neutrinos propagate through the Earth, and a fraction are detected by IceCube.
Moreover, a subdominant flux of detected neutrinos, in addition to those made in the atmosphere, are astrophysical in origin.
These neutrinos are diffuse in direction and fall less steeply in energy, as approximately $\sim E^{-2.5}$; they can also be electron, muon, or tau-flavored.
Cosmic-ray muons --- the dominant background --- cannot traverse the planet from upgoing directions, making these events clean neutrino candidates.
In this analysis, we assume a small, fixed, background of misreconstructed atmospheric muons, which is supported by the data and Monte Carlo comparisons in Ref.~\cite{IceCubeCollaboration:2024dxk}, assuming the PREM density profile.

A key signature of muon neutrino charged-current interactions in the detector is the \textit{track} morphology: the long, linear pattern of Cherenkov light emitted by the resulting muon.
While electron and tau neutrinos may also yield track-like signatures, their expected contributions, after accounting for the relevant branching fractions and fluxes, are negligible compared to those from muon-neutrino interactions.
Upgoing tracks, corresponding to neutrinos which come from below the horizon, are particularly suitable for Earth tomography, as they correspond to neutrinos that have traversed the Earth and whose flux is attenuated in a way that depends on the column density along their path.
We further subdivide the track morphology into two categories: \textit{starting} and \textit{throughgoing} tracks.
Starting tracks arise from neutrino interactions occurring within the instrumented volume of IceCube, whereas throughgoing tracks are produced by neutrino interactions outside the detector, with the resulting muon traversing the array.
We reconstruct the neutrino energy, $E_\nu$ and zenith angle, $\theta$, for all tracks, but the energy reconstruction is more accurate~\cite{IceCubeCollaboration:2024dxk} for starting tracks owing to the interaction occurring directly within the instrumented volume.
The dataset contains $368{,}071$ tracks (see Ref.~\cite{IceCubeCollaboration:2024dxk}), collected over a 10.7-year period from 2011-2022.

To compare the observed energy and zenith distributions of tracks with predictions from a given Earth model, we parametrize the Earth's density profile as a set of uniform-density concentric shells, approximating the profile shown in the top right of Fig.~\ref{fig:schematic}.
Throughout, ``density'' refers to mass density, $\rho$; neutrinos probe this via the nucleon number density $n_N = \rho/m_N$, assuming a fixed nuclear composition.
These shells are assigned uniform, nonnegative priors.
We consider a five-shell configuration, following Ref.~\cite{Donini:2018tsg}, as well as a finer-grained eight-shell extension described in the Supplementary Information. 
Some radial boundaries of these shells are set at strategic transition points, like the core-mantle boundary, but are otherwise set to split the Earth into shells of the same thickness.
For a given density profile and nuisance parameter values, we predict the expected event distributions using Monte Carlo simulations of the IceCube detector, which model its acceptance and systematic response.
The predicted distributions consist of a mixture of components: the conventional atmospheric neutrino flux (arising from pion and kaon decays in cosmic-ray air showers), the diffuse astrophysical neutrino flux modelled as a broken power law, and a subdominant prompt atmospheric neutrino contribution from charmed meson decays.
Each component is reweighted according to the current density and nuisance parameters, and the resulting distributions are computed as functions of $\cos\theta$ and $E_\nu$, separately for starting and throughgoing tracks.
We then perform a joint fit for the shell densities and nuisance parameters using a binned Poisson likelihood, with nuisance parameters varied simultaneously to account for flux, cross-section, detector, and ice uncertainties, following an approach similar to that of Refs.~\cite{IceCubeCollaboration:2024nle,IceCubeCollaboration:2024dxk}.
The joint posterior distributions of all parameters are sampled with a Markov Chain Monte Carlo algorithm~\cite{Goodman:2010dyf} implemented in the \texttt{emcee} software~\cite{emcee2013}.

\section{Results}\label{sec:results}

\begin{figure}
    \centering
    \includegraphics[width=0.7\linewidth]{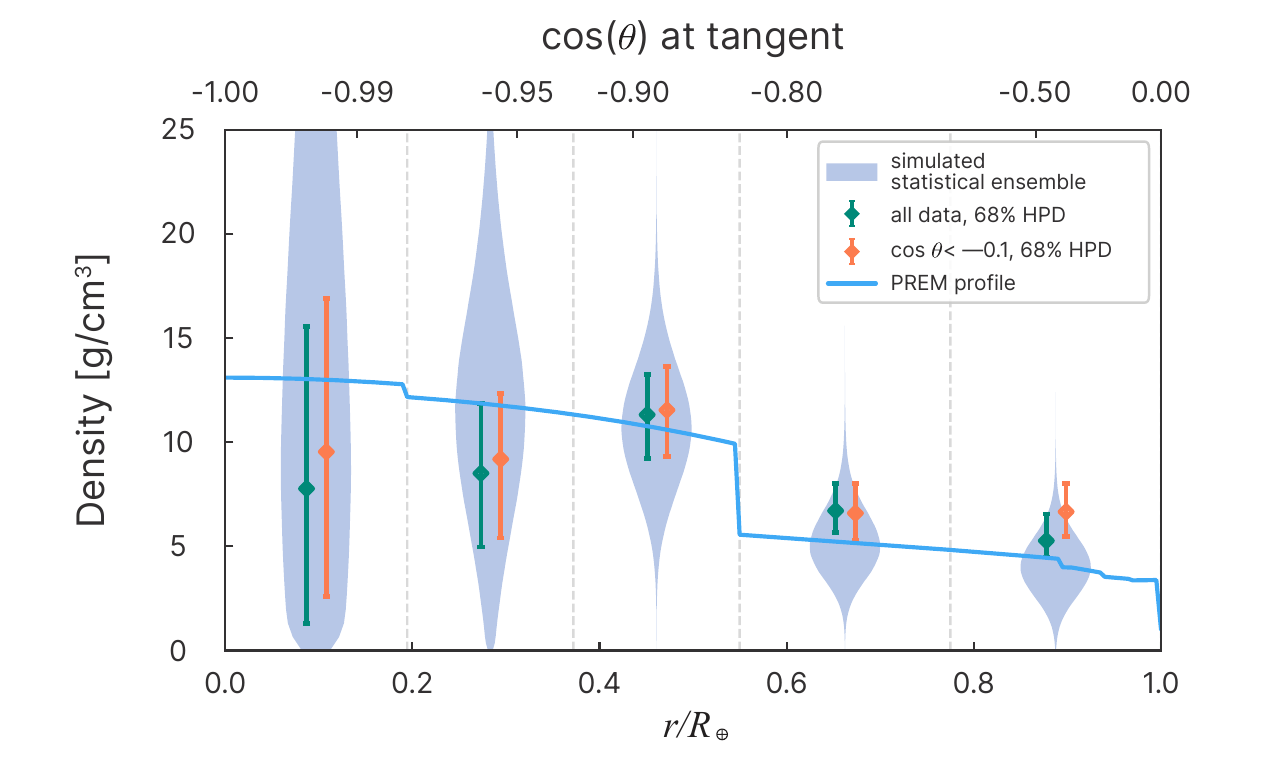}
    \caption{\textbf{Measured radial density profile of the Earth from high-energy neutrino attenuation (five-shell parametrization).}
    The Earth is parameterized as five uniform-density concentric shells (vertical dashed grey lines), and the density in each shell is inferred from the energy- and zenith-dependent attenuation of upgoing atmospheric muon neutrinos observed by IceCube, assuming the CSMS neutrino cross section in the Standard Model \cite{Cooper-Sarkar:2011jtt}.
    We show shell densities at the shell midpoints versus normalized radius, $r/R_\oplus$; points indicate posterior modes and error bars give 68\% highest-posterior-density (HPD) credible intervals.
    Results are shown for the blind fit excluding near-horizontal events ($-1.0 < \cos\theta < -0.1$) and for a fit with all data.
    The violin plots show the distribution expected from statistical fluctuations alone, obtained by repeating the full inference on pseudo-data generated from a nominal model with PREM-averaged densities in each shell; the solid blue curve shows the PREM profile for comparison.
    The upper axis maps radial depth to arrival direction via the $\cos\theta$ of trajectories tangent to radius $r$, giving a sense of which $\cos\theta$ ranges map to which corresponding radius ranges.
    The fraction of the marginal posterior lying below the PREM-normalised value of unity for shells 1 through 5 is $61\%$, $83\%$, $38\%$, $7.2\%$, $9.2\%$ (all data) and $61\%$, $79\%$, $31\%$, $9.7\%$, $1.7\%$ (excl.\ near-horizontal events).}
    \label{fig:5B_profile}
\end{figure}

Our main result is the inferred density profile from the five-shell parametrization, shown in Fig.~\ref{fig:5B_profile}.
We perform a blind analysis, meaning that we do not look at the density posterior results until the analysis procedure is finalized and all data selection steps are fixed based on studies with simulation only; the purpose of this is to avoid any bias in analysis design based on the observed data.
The inference was performed in stages, with the first stage being a blind fit to data excluding near-horizontal events, defined as events whose reconstructed zenith angle satisfies $-1.0 < \cos\theta < -0.1$.
The subsequent stage is a fit to all data, including near-horizontal events, satisfying $-1.0 < \cos\theta < 0.0$, but otherwise the same fit.
The reason for this two-stage process is that the near-horizontal region corresponds to trajectories with the smallest column depths, where detector and atmospheric effects have a higher chance of being mismodelled; a fit with and without this region allows us to check the stability of the results and to see if there is significant evidence of such mismodelling.
Seeing none, we present both fits for comparison: we report the 68\% highest-posterior-density (HPD) region of the density posterior for all five shells for both cases.
Fig.~\ref{fig:5B_profile} also shows a violin plot of expected statistical fluctuations, constructed by fitting to 50 realizations of the model given the nominal density profile (based on the average PREM density in each shell) and superimposing the resulting posteriors (with the violins representing the KDE of the superposition).
This indicates the size and frequency of statistical fluctuations given a model consistent with the PREM; the measurements, well within these ranges, indicate good agreement with the PREM model.
Table~\ref{tab:density_results} also shows the numerical values of the density posterior modes and the 68\% HPD ranges, normalized to the average PREM density in that layer; this PREM density is also shown in the table in parentheses.
As a straightforward extension providing finer radial granularity, we repeat the analysis with an eight-shell parametrization; this result is shown in Supplementary Materials, Fig.~\ref{fig:8B_profile}.


\begin{table}
    \centering
    \footnotesize
    \caption{\textbf{Posterior constraints on the Earth shell densities.}
    Posterior modes and 68\% HPD credible intervals are shown for the five density shells, for both fits with and without the near-horizontal region.
    Values are expressed as multiplicative factors relative to the nominal density of each shell derived from the Preliminary Reference Earth Model (PREM); the PREM densities used for normalization are listed in parentheses.
    Shell 1 corresponds to the inner core, shells 2 and 3 correspond to the outer core, and shells 4 and 5 correspond to the mantle and crust.}
    \label{tab:density_results}
    \begin{tabular}{lcccccc}
    \toprule
    & \multicolumn{3}{c}{Excl.\ horizon ($\cos\theta < -0.1$)} & \multicolumn{3}{c}{All data} \\
    \cmidrule(lr){2-4}\cmidrule(lr){5-7}
    Posterior & \quad Mode \quad & \quad Lower (68\% HPD) \quad & \quad Upper (68\% HPD) \quad & \quad Mode \quad & \quad Lower (68\% HPD) \quad & \quad Upper (68\% HPD) \quad \\
    \midrule
    Shell density 1 / ($13\,\mathrm{g\,cm^{-3}}$) & 0.73 & 0.20 & 1.3 & 0.60 & 0.099 & 1.2 \\
    Shell density 2 / ($12\,\mathrm{g\,cm^{-3}}$) & 0.78 & 0.46 & 1.0 & 0.72 & 0.42 & 1.0 \\
    Shell density 3 / ($11\,\mathrm{g\,cm^{-3}}$) & 1.1  & 0.87 & 1.3 & 1.1  & 0.86 & 1.2 \\
    Shell density 4 / ($5.2\,\mathrm{g\,cm^{-3}}$) & 1.3 & 1.0  & 1.5 & 1.3  & 1.1  & 1.5 \\
    Shell density 5 / ($4.1\,\mathrm{g\,cm^{-3}}$) & 1.6 & 1.3  & 1.9 & 1.3  & 1.1  & 1.6 \\
    \bottomrule
    \end{tabular}
\end{table}

\begin{figure}
    \centering
    \includegraphics[width=0.6\linewidth]{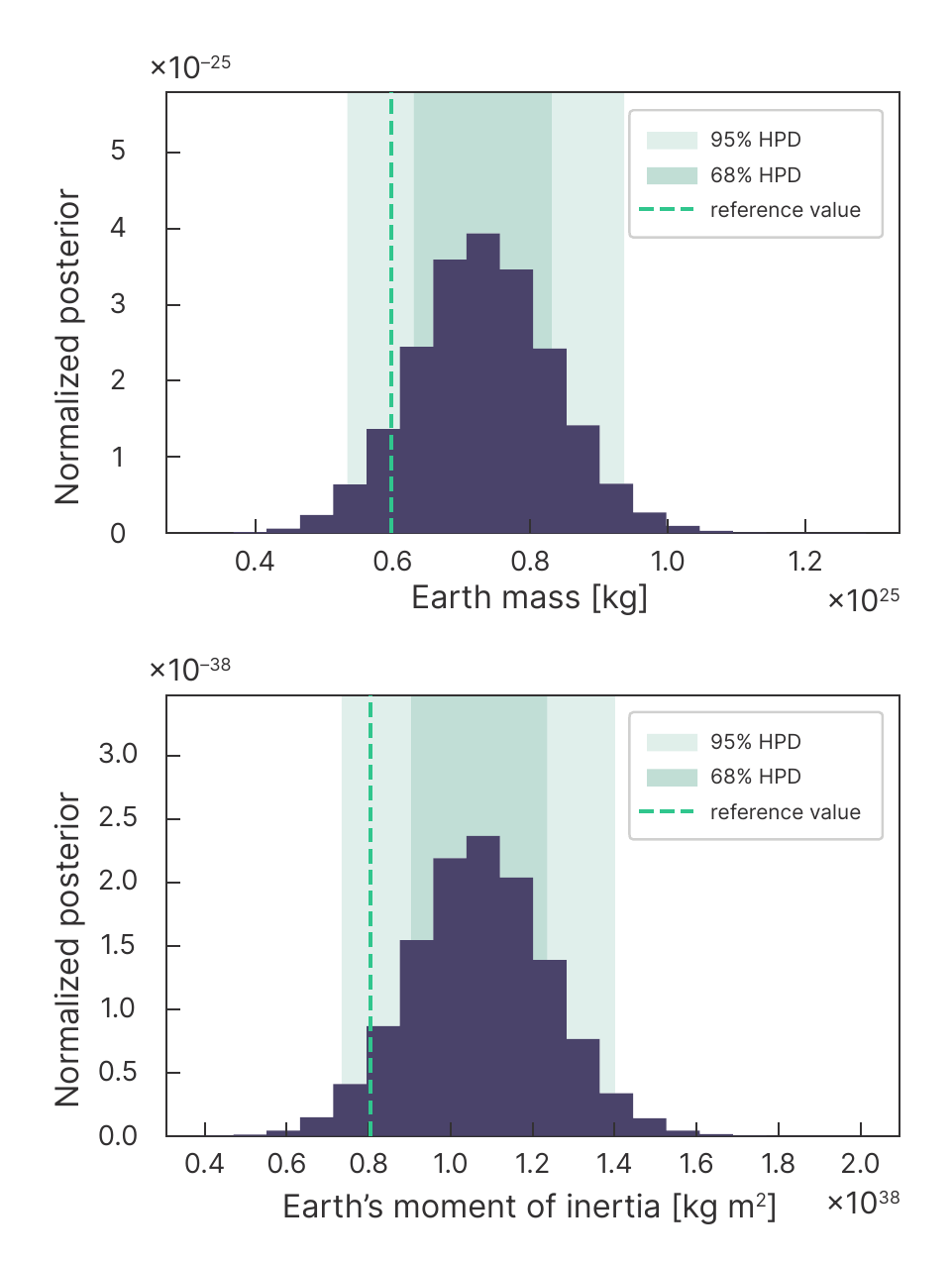}
    \caption{\textbf{The posterior distributions of the Earth's mass and moment of inertia.} 
    These are inferred from the measured density profile. 
    In the top panel, we show the mass posterior, derived from our density profile measurement in Fig.~\ref{fig:5B_profile}, and a vertical dashed line showing the reference value, the mass as reported in Ref.~\cite{2011CeMDA.110..293L}, measured through gravitational means. 
    About $9\%$ of the posterior density lies below this reference value.
    In the bottom panel, we show the equivalent figure for the moment of inertia, assuming the Earth is spherically-symmetric, with a variable radial density profile.
    The vertical dashed line shows the reference value as obtained from the PREM model~\cite{DZIEWONSKI1981297}.
    About $6\%$ of the posterior density is below this value.}
    \label{fig:5B_mass_post}
\end{figure}

By inferring the density profile, we can also infer derived quantities, such as the mass and moment of inertia of the Earth. 
Fig.~\ref{fig:5B_mass_post} (top) shows the mass, $M$, posterior of the Earth derived from our density measurement, obtained by computing the mass predicted by each density profile sample, assuming the Earth is a perfect sphere.
We find a posterior mode of $M = 7.25 \times 10^{24}\,\mathrm{kg}$, with a 68\% HPD interval of $[6.31,\, 8.31] \times 10^{24}\,\mathrm{kg}$.
This may be compared to the gravitational mass of the Earth, $5.97 \times 10^{24}\,\mathrm{kg}$~\cite{2011CeMDA.110..293L}, indicated by the vertical dashed line; while our posterior mode lies above this value, the gravitational mass is fully contained within our 95\% HPD interval, with approximately 9\% of the posterior distribution lying below it.
As a measure of distance from the vacuum Earth hypothesis (zero Earth mass as probed by neutrinos), the value of zero is over $5\sigma$ away from the mode of the distribution, where $\sigma$ is the standard deviation of the (approximately Gaussian) posterior.

In a similar way, we obtain the polar moment of inertia, $I$, posterior distribution from the same density samples (assuming spherical symmetry), which is shown in Fig.~\ref{fig:5B_mass_post} (bottom).
We find a posterior mode of $I = 1.05 \times 10^{38}\,\mathrm{kg}\,\mathrm{m}^2$, with a 68\% HPD interval of $[9.04 \times 10^{37},\, 1.24 \times 10^{38}]\,\mathrm{kg}\,\mathrm{m}^2$, to be compared with the reference value of $8.01 \times 10^{37}\,\mathrm{kg}\,\mathrm{m}^2$~\cite{DZIEWONSKI1981297}.
This reference value is contained within our 95\% HPD interval, with approximately 6\% of the posterior distribution lying below it, and the posterior mode lies above the reference value, consistent with the mass result.
Combining the posterior mode of the moment of inertia with the measured mass of the Earth $M = 7.25 \times 10^{24}\,\mathrm{kg}$, and its mean radius of $R_\oplus = 6371\,\mathrm{km}$, we obtain the normalized moment of inertia $I/(MR_\oplus^2) \approx 0.357$, comparable to the reference value used in the PREM, $0.331$~\cite{DZIEWONSKI1981297}.

\section{Discussion}

We have presented the most precise weak-interaction measurement of the Earth's interior to date using high-energy atmospheric and astrophysical neutrinos observed by IceCube. 
The observed attenuation pattern of upgoing events encodes the column density along different chords through the planet, allowing us to infer the Earth's radial density profile in a concentric-shell parametrization. The same posterior samples yield the Earth's mass and polar moment of inertia under spherical symmetry.

Our first unblinded result excludes near-horizontal events ($-1.0 < \cos\theta < -0.1$), where trajectories have the smallest column depths and detector and atmospheric effects can compete with attenuation.
The subsequent fit, using all available data by comparison ($-1.0 < \cos\theta < 0.0$), shows that the inferred density profile is generally stable under this choice.
However, in including all data, the density posterior for the outermost shell (shell 5) posterior mode lies closer to the PREM-normalised value of unity (mode $1.3$ versus $1.6$, with the 68\% HPD interval $[1.1, 1.6]$ more closely encompassing unity than $[1.3, 1.9]$; see Table~\ref{tab:density_results}), suggesting that the near-horizontal events contribute constructively, albeit not significantly, to the overall constraint.
The $\cos\theta = -0.1$ cut removes $17.1\%$ of the total event sample available ($62{,}887$ of $368{,}071$ events).

The derived mass and moment-of-inertia posteriors are consistent with gravitational determinations and with the seismologically-derived PREM profile within the inferred credible regions. 
The mild upward shift of the posterior modes relative to reference values is compatible with the systematic uncertainties due to cross section, flux, and detector/ice effects, as well as statistical variations as exhibited by the violin plots in Figs.~\ref{fig:5B_profile} and \ref{fig:8B_profile}.
These uncertainties are expected to be reduced by improved calibration and external constraints (for instance, from the IceCube Upgrade~\cite{Ishihara:2019aao}).

Looking forward, larger next-generation neutrino telescopes~\cite{IceCube-Gen2:2020qha} and those based in water~\cite{KM3Net:2016zxf}, not ice, will increase event statistics above $\sim10\,\mathrm{TeV}$ and reduce detector and ice uncertainties, strengthening the attenuation signal that underpins tomography. 
In parallel, improved constraints on the neutrino-nucleon cross section and atmospheric flux modelling will translate directly into tighter posteriors. 
Together, these advances can extend neutrino attenuation tomography from a complementary cross-check of Earth structure to a precision weak-interaction-based probe of the Earth's interior.


\bibliography{main.bib}

\begin{thebibliography}{48}%
\makeatletter
\providecommand \@ifxundefined [1]{%
 \@ifx{#1\undefined}
}%
\providecommand \@ifnum [1]{%
 \ifnum #1\expandafter \@firstoftwo
 \else \expandafter \@secondoftwo
 \fi
}%
\providecommand \@ifx [1]{%
 \ifx #1\expandafter \@firstoftwo
 \else \expandafter \@secondoftwo
 \fi
}%
\providecommand \natexlab [1]{#1}%
\providecommand \enquote  [1]{``#1''}%
\providecommand \bibnamefont  [1]{#1}%
\providecommand \bibfnamefont [1]{#1}%
\providecommand \citenamefont [1]{#1}%
\providecommand \href@noop [0]{\@secondoftwo}%
\providecommand \href [0]{\begingroup \@sanitize@url \@href}%
\providecommand \@href[1]{\@@startlink{#1}\@@href}%
\providecommand \@@href[1]{\endgroup#1\@@endlink}%
\providecommand \@sanitize@url [0]{\catcode `\\12\catcode `\$12\catcode `\&12\catcode `\#12\catcode `\^12\catcode `\_12\catcode `\%12\relax}%
\providecommand \@@startlink[1]{}%
\providecommand \@@endlink[0]{}%
\providecommand \url  [0]{\begingroup\@sanitize@url \@url }%
\providecommand \@url [1]{\endgroup\@href {#1}{\urlprefix }}%
\providecommand \urlprefix  [0]{URL }%
\providecommand \Eprint [0]{\href }%
\providecommand \doibase [0]{https://doi.org/}%
\providecommand \selectlanguage [0]{\@gobble}%
\providecommand \bibinfo  [0]{\@secondoftwo}%
\providecommand \bibfield  [0]{\@secondoftwo}%
\providecommand \translation [1]{[#1]}%
\providecommand \BibitemOpen [0]{}%
\providecommand \bibitemStop [0]{}%
\providecommand \bibitemNoStop [0]{.\EOS\space}%
\providecommand \EOS [0]{\spacefactor3000\relax}%
\providecommand \BibitemShut  [1]{\csname bibitem#1\endcsname}%
\let\auto@bib@innerbib\@empty
\bibitem [{\citenamefont {Williamson}\ and\ \citenamefont {Adams}(1923)}]{williamson1923density}%
  \BibitemOpen
  \bibfield  {author} {\bibinfo {author} {\bibfnamefont {E.~D.}\ \bibnamefont {Williamson}}\ and\ \bibinfo {author} {\bibfnamefont {L.~H.}\ \bibnamefont {Adams}},\ }\href@noop {} {\bibfield  {journal} {\bibinfo  {journal} {Journal of the Washington Academy of Sciences}\ }\textbf {\bibinfo {volume} {13}},\ \bibinfo {pages} {413} (\bibinfo {year} {1923})}\BibitemShut {NoStop}%
\bibitem [{\citenamefont {Clark~Jr}\ and\ \citenamefont {Ringwood}(1964)}]{clark1964density}%
  \BibitemOpen
  \bibfield  {author} {\bibinfo {author} {\bibfnamefont {S.~P.}\ \bibnamefont {Clark~Jr}}\ and\ \bibinfo {author} {\bibfnamefont {A.~E.}\ \bibnamefont {Ringwood}},\ }\href@noop {} {\bibfield  {journal} {\bibinfo  {journal} {Reviews of Geophysics}\ }\textbf {\bibinfo {volume} {2}},\ \bibinfo {pages} {35} (\bibinfo {year} {1964})}\BibitemShut {NoStop}%
\bibitem [{\citenamefont {Dziewonski}\ and\ \citenamefont {Anderson}(1981)}]{DZIEWONSKI1981297}%
  \BibitemOpen
  \bibfield  {author} {\bibinfo {author} {\bibfnamefont {A.~M.}\ \bibnamefont {Dziewonski}}\ and\ \bibinfo {author} {\bibfnamefont {D.~L.}\ \bibnamefont {Anderson}},\ }\href {https://doi.org/https://doi.org/10.1016/0031-9201(81)90046-7} {\bibfield  {journal} {\bibinfo  {journal} {Physics of the Earth and Planetary Interiors}\ }\textbf {\bibinfo {volume} {25}},\ \bibinfo {pages} {297} (\bibinfo {year} {1981})}\BibitemShut {NoStop}%
\bibitem [{\citenamefont {Gilbert}\ and\ \citenamefont {Dziewonski}(1975)}]{Gilbert1975}%
  \BibitemOpen
  \bibfield  {author} {\bibinfo {author} {\bibfnamefont {F.}~\bibnamefont {Gilbert}}\ and\ \bibinfo {author} {\bibfnamefont {A.~M.}\ \bibnamefont {Dziewonski}},\ }\href {https://doi.org/10.1098/rsta.1975.0025} {\bibfield  {journal} {\bibinfo  {journal} {Philosophical Transactions of the Royal Society of London, Series A: Mathematical and Physical Sciences}\ }\textbf {\bibinfo {volume} {278}},\ \bibinfo {pages} {187} (\bibinfo {year} {1975})}\BibitemShut {NoStop}%
\bibitem [{\citenamefont {Kennett}\ \emph {et~al.}(1995)\citenamefont {Kennett}, \citenamefont {Engdahl},\ and\ \citenamefont {Buland}}]{AK135:1995}%
  \BibitemOpen
  \bibfield  {author} {\bibinfo {author} {\bibfnamefont {B.~L.~N.}\ \bibnamefont {Kennett}}, \bibinfo {author} {\bibfnamefont {E.~R.}\ \bibnamefont {Engdahl}},\ and\ \bibinfo {author} {\bibfnamefont {R.}~\bibnamefont {Buland}},\ }\href {https://doi.org/10.1111/j.1365-246X.1995.tb03540.x} {\bibfield  {journal} {\bibinfo  {journal} {Geophysical Journal International}\ }\textbf {\bibinfo {volume} {122}},\ \bibinfo {pages} {108} (\bibinfo {year} {1995})}\BibitemShut {NoStop}%
\bibitem [{\citenamefont {Garcia}\ \emph {et~al.}(2020)\citenamefont {Garcia}, \citenamefont {Gauld}, \citenamefont {Heijboer},\ and\ \citenamefont {Rojo}}]{Garcia:2020jwr}%
  \BibitemOpen
  \bibfield  {author} {\bibinfo {author} {\bibfnamefont {A.}~\bibnamefont {Garcia}}, \bibinfo {author} {\bibfnamefont {R.}~\bibnamefont {Gauld}}, \bibinfo {author} {\bibfnamefont {A.}~\bibnamefont {Heijboer}},\ and\ \bibinfo {author} {\bibfnamefont {J.}~\bibnamefont {Rojo}},\ }\href {https://doi.org/10.1088/1475-7516/2020/09/025} {\bibfield  {journal} {\bibinfo  {journal} {JCAP}\ }\textbf {\bibinfo {volume} {09}},\ \bibinfo {pages} {025}},\ \Eprint {https://arxiv.org/abs/2004.04756} {arXiv:2004.04756 [hep-ph]} \BibitemShut {NoStop}%
\bibitem [{\citenamefont {Reno}(2023)}]{Reno:2023sdm}%
  \BibitemOpen
  \bibfield  {author} {\bibinfo {author} {\bibfnamefont {M.~H.}\ \bibnamefont {Reno}},\ }\href {https://doi.org/10.1146/annurev-nucl-111422-040200} {\bibfield  {journal} {\bibinfo  {journal} {Ann. Rev. Nucl. Part. Sci.}\ }\textbf {\bibinfo {volume} {73}},\ \bibinfo {pages} {181} (\bibinfo {year} {2023})}\BibitemShut {NoStop}%
\bibitem [{\citenamefont {Gonzalez-Garcia}\ \emph {et~al.}(2008)\citenamefont {Gonzalez-Garcia}, \citenamefont {Halzen}, \citenamefont {Maltoni},\ and\ \citenamefont {Tanaka}}]{Gonzalez-Garcia:2007wfs}%
  \BibitemOpen
  \bibfield  {author} {\bibinfo {author} {\bibfnamefont {M.~C.}\ \bibnamefont {Gonzalez-Garcia}}, \bibinfo {author} {\bibfnamefont {F.}~\bibnamefont {Halzen}}, \bibinfo {author} {\bibfnamefont {M.}~\bibnamefont {Maltoni}},\ and\ \bibinfo {author} {\bibfnamefont {H.~K.~M.}\ \bibnamefont {Tanaka}},\ }\href {https://doi.org/10.1103/PhysRevLett.100.061802} {\bibfield  {journal} {\bibinfo  {journal} {Phys. Rev. Lett.}\ }\textbf {\bibinfo {volume} {100}},\ \bibinfo {pages} {061802} (\bibinfo {year} {2008})},\ \Eprint {https://arxiv.org/abs/0711.0745} {arXiv:0711.0745 [hep-ph]} \BibitemShut {NoStop}%
\bibitem [{\citenamefont {Vincent}\ \emph {et~al.}(2017)\citenamefont {Vincent}, \citenamefont {Arg{\"u}elles},\ and\ \citenamefont {Kheirandish}}]{Vincent:2017svp}%
  \BibitemOpen
  \bibfield  {author} {\bibinfo {author} {\bibfnamefont {A.~C.}\ \bibnamefont {Vincent}}, \bibinfo {author} {\bibfnamefont {C.~A.}\ \bibnamefont {Arg{\"u}elles}},\ and\ \bibinfo {author} {\bibfnamefont {A.}~\bibnamefont {Kheirandish}},\ }\href {https://doi.org/10.1088/1475-7516/2017/11/012} {\bibfield  {journal} {\bibinfo  {journal} {JCAP}\ }\textbf {\bibinfo {volume} {11}},\ \bibinfo {pages} {012}},\ \Eprint {https://arxiv.org/abs/1706.09895} {arXiv:1706.09895 [hep-ph]} \BibitemShut {NoStop}%
\bibitem [{\citenamefont {Abbasi}\ \emph {et~al.}(2021{\natexlab{a}})\citenamefont {Abbasi} \emph {et~al.}}]{IceCube:2020rnc}%
  \BibitemOpen
  \bibfield  {author} {\bibinfo {author} {\bibfnamefont {R.}~\bibnamefont {Abbasi}} \emph {et~al.} (\bibinfo {collaboration} {IceCube}),\ }\href {https://doi.org/10.1103/PhysRevD.104.022001} {\bibfield  {journal} {\bibinfo  {journal} {Phys. Rev. D}\ }\textbf {\bibinfo {volume} {104}},\ \bibinfo {pages} {022001} (\bibinfo {year} {2021}{\natexlab{a}})},\ \Eprint {https://arxiv.org/abs/2011.03560} {arXiv:2011.03560 [hep-ex]} \BibitemShut {NoStop}%
\bibitem [{\citenamefont {Aartsen}\ \emph {et~al.}(2017)\citenamefont {Aartsen} \emph {et~al.}}]{IceCube:2016zyt}%
  \BibitemOpen
  \bibfield  {author} {\bibinfo {author} {\bibfnamefont {M.~G.}\ \bibnamefont {Aartsen}} \emph {et~al.} (\bibinfo {collaboration} {IceCube}),\ }\href {https://doi.org/10.1088/1748-0221/12/03/P03012} {\bibfield  {journal} {\bibinfo  {journal} {JINST}\ }\textbf {\bibinfo {volume} {12}}\bibfield  {number} {\bibinfo  {number} { (03)},\ \bibinfo {pages} {P03012}},\ }\bibinfo {note} {[Erratum: JINST 19, E05001 (2024)]},\ \Eprint {https://arxiv.org/abs/1612.05093} {arXiv:1612.05093 [astro-ph.IM]} \BibitemShut {NoStop}%
\bibitem [{\citenamefont {Placci}\ and\ \citenamefont {Zavattini}(1973)}]{Placci:1973yvk}%
  \BibitemOpen
  \bibfield  {author} {\bibinfo {author} {\bibfnamefont {A.}~\bibnamefont {Placci}}\ and\ \bibinfo {author} {\bibfnamefont {E.}~\bibnamefont {Zavattini}},\ }\href@noop {} {\bibfield  {journal} {\bibinfo  {journal} {CERN Internal Report}\ } (\bibinfo {year} {1973})}\BibitemShut {NoStop}%
\bibitem [{\citenamefont {Volkova}\ and\ \citenamefont {Zatsepin}(1974)}]{Volkova:1974xa}%
  \BibitemOpen
  \bibfield  {author} {\bibinfo {author} {\bibfnamefont {L.~V.}\ \bibnamefont {Volkova}}\ and\ \bibinfo {author} {\bibfnamefont {G.~T.}\ \bibnamefont {Zatsepin}},\ }\href@noop {} {\bibfield  {journal} {\bibinfo  {journal} {Izv. Akad. Nauk Ser. Fiz.}\ }\textbf {\bibinfo {volume} {38N5}},\ \bibinfo {pages} {1060} (\bibinfo {year} {1974})}\BibitemShut {NoStop}%
\bibitem [{\citenamefont {Reynoso}\ and\ \citenamefont {Sampayo}(2004)}]{Reynoso:2004dt}%
  \BibitemOpen
  \bibfield  {author} {\bibinfo {author} {\bibfnamefont {M.~M.}\ \bibnamefont {Reynoso}}\ and\ \bibinfo {author} {\bibfnamefont {O.~A.}\ \bibnamefont {Sampayo}},\ }\href {https://doi.org/10.1016/j.astropartphys.2004.01.003} {\bibfield  {journal} {\bibinfo  {journal} {Astropart. Phys.}\ }\textbf {\bibinfo {volume} {21}},\ \bibinfo {pages} {315} (\bibinfo {year} {2004})},\ \Eprint {https://arxiv.org/abs/hep-ph/0401102} {arXiv:hep-ph/0401102} \BibitemShut {NoStop}%
\bibitem [{\citenamefont {{Hoshina}}\ and\ \citenamefont {{Tanaka}}(2012)}]{2012EGUGA}%
  \BibitemOpen
  \bibfield  {author} {\bibinfo {author} {\bibfnamefont {K.}~\bibnamefont {{Hoshina}}}\ and\ \bibinfo {author} {\bibfnamefont {H.~K.~M.}\ \bibnamefont {{Tanaka}}},\ }in\ \href@noop {} {\emph {\bibinfo {booktitle} {EGU General Assembly Conference Abstracts}}},\ \bibinfo {series and number} {EGU General Assembly Conference Abstracts}\ (\bibinfo {year} {2012})\ p.\ \bibinfo {pages} {3246}\BibitemShut {NoStop}%
\bibitem [{\citenamefont {Donini}\ \emph {et~al.}(2019)\citenamefont {Donini}, \citenamefont {Palomares-Ruiz},\ and\ \citenamefont {Salvado}}]{Donini:2018tsg}%
  \BibitemOpen
  \bibfield  {author} {\bibinfo {author} {\bibfnamefont {A.}~\bibnamefont {Donini}}, \bibinfo {author} {\bibfnamefont {S.}~\bibnamefont {Palomares-Ruiz}},\ and\ \bibinfo {author} {\bibfnamefont {J.}~\bibnamefont {Salvado}},\ }\href {https://doi.org/10.1038/s41567-018-0319-1} {\bibfield  {journal} {\bibinfo  {journal} {Nature Phys.}\ }\textbf {\bibinfo {volume} {15}},\ \bibinfo {pages} {37} (\bibinfo {year} {2019})},\ \Eprint {https://arxiv.org/abs/1803.05901} {arXiv:1803.05901 [hep-ph]} \BibitemShut {NoStop}%
\bibitem [{\citenamefont {Abbasi}\ \emph {et~al.}(2010)\citenamefont {Abbasi} \emph {et~al.}}]{IceCube:2010dpc}%
  \BibitemOpen
  \bibfield  {author} {\bibinfo {author} {\bibfnamefont {R.}~\bibnamefont {Abbasi}} \emph {et~al.} (\bibinfo {collaboration} {IceCube}),\ }\href {https://doi.org/10.1016/j.nima.2010.03.102} {\bibfield  {journal} {\bibinfo  {journal} {Nucl. Instrum. Meth. A}\ }\textbf {\bibinfo {volume} {618}},\ \bibinfo {pages} {139} (\bibinfo {year} {2010})},\ \Eprint {https://arxiv.org/abs/1002.2442} {arXiv:1002.2442 [astro-ph.IM]} \BibitemShut {NoStop}%
\bibitem [{\citenamefont {Navas}\ \emph {et~al.}(2024)\citenamefont {Navas} \emph {et~al.}}]{ParticleDataGroup:2024cfk}%
  \BibitemOpen
  \bibfield  {author} {\bibinfo {author} {\bibfnamefont {S.}~\bibnamefont {Navas}} \emph {et~al.} (\bibinfo {collaboration} {Particle Data Group}),\ }\href {https://doi.org/10.1103/PhysRevD.110.030001} {\bibfield  {journal} {\bibinfo  {journal} {Phys. Rev. D}\ }\textbf {\bibinfo {volume} {110}},\ \bibinfo {pages} {030001} (\bibinfo {year} {2024})}\BibitemShut {NoStop}%
\bibitem [{\citenamefont {Devenish}\ and\ \citenamefont {Cooper-Sarkar}(2004)}]{Devenish:2004pb}%
  \BibitemOpen
  \bibfield  {author} {\bibinfo {author} {\bibfnamefont {R.}~\bibnamefont {Devenish}}\ and\ \bibinfo {author} {\bibfnamefont {A.}~\bibnamefont {Cooper-Sarkar}},\ }\href {https://doi.org/10.1093/acprof:oso/9780198506713.001.0001} {\emph {\bibinfo {title} {{Deep inelastic scattering}}}}\ (\bibinfo  {publisher} {Oxford University Press},\ \bibinfo {year} {2004})\BibitemShut {NoStop}%
\bibitem [{\citenamefont {Dokshitzer}(1977)}]{Dokshitzer:1977sg}%
  \BibitemOpen
  \bibfield  {author} {\bibinfo {author} {\bibfnamefont {Y.~L.}\ \bibnamefont {Dokshitzer}},\ }\href@noop {} {\bibfield  {journal} {\bibinfo  {journal} {Sov. Phys. JETP}\ }\textbf {\bibinfo {volume} {46}},\ \bibinfo {pages} {641} (\bibinfo {year} {1977})}\BibitemShut {NoStop}%
\bibitem [{\citenamefont {Gribov}\ and\ \citenamefont {Lipatov}(1972{\natexlab{a}})}]{Gribov:1972ri}%
  \BibitemOpen
  \bibfield  {author} {\bibinfo {author} {\bibfnamefont {V.~N.}\ \bibnamefont {Gribov}}\ and\ \bibinfo {author} {\bibfnamefont {L.~N.}\ \bibnamefont {Lipatov}},\ }\href@noop {} {\bibfield  {journal} {\bibinfo  {journal} {Sov. J. Nucl. Phys.}\ }\textbf {\bibinfo {volume} {15}},\ \bibinfo {pages} {438} (\bibinfo {year} {1972}{\natexlab{a}})}\BibitemShut {NoStop}%
\bibitem [{\citenamefont {Gribov}\ and\ \citenamefont {Lipatov}(1972{\natexlab{b}})}]{Gribov:1972rt}%
  \BibitemOpen
  \bibfield  {author} {\bibinfo {author} {\bibfnamefont {V.~N.}\ \bibnamefont {Gribov}}\ and\ \bibinfo {author} {\bibfnamefont {L.~N.}\ \bibnamefont {Lipatov}},\ }\href@noop {} {\bibfield  {journal} {\bibinfo  {journal} {Sov. J. Nucl. Phys.}\ }\textbf {\bibinfo {volume} {15}},\ \bibinfo {pages} {675} (\bibinfo {year} {1972}{\natexlab{b}})}\BibitemShut {NoStop}%
\bibitem [{\citenamefont {Altarelli}\ and\ \citenamefont {Parisi}(1977)}]{Altarelli:1977zs}%
  \BibitemOpen
  \bibfield  {author} {\bibinfo {author} {\bibfnamefont {G.}~\bibnamefont {Altarelli}}\ and\ \bibinfo {author} {\bibfnamefont {G.}~\bibnamefont {Parisi}},\ }\href {https://doi.org/10.1016/0550-3213(77)90384-4} {\bibfield  {journal} {\bibinfo  {journal} {Nucl. Phys. B}\ }\textbf {\bibinfo {volume} {126}},\ \bibinfo {pages} {298} (\bibinfo {year} {1977})}\BibitemShut {NoStop}%
\bibitem [{\citenamefont {Aaron}\ \emph {et~al.}(2010)\citenamefont {Aaron} \emph {et~al.}}]{H1:2009pze}%
  \BibitemOpen
  \bibfield  {author} {\bibinfo {author} {\bibfnamefont {F.~D.}\ \bibnamefont {Aaron}} \emph {et~al.} (\bibinfo {collaboration} {H1, ZEUS}),\ }\href {https://doi.org/10.1007/JHEP01(2010)109} {\bibfield  {journal} {\bibinfo  {journal} {JHEP}\ }\textbf {\bibinfo {volume} {01}},\ \bibinfo {pages} {109}},\ \Eprint {https://arxiv.org/abs/0911.0884} {arXiv:0911.0884 [hep-ex]} \BibitemShut {NoStop}%
\bibitem [{\citenamefont {Abramowicz}\ \emph {et~al.}(2015)\citenamefont {Abramowicz} \emph {et~al.}}]{H1:2015ubc}%
  \BibitemOpen
  \bibfield  {author} {\bibinfo {author} {\bibfnamefont {H.}~\bibnamefont {Abramowicz}} \emph {et~al.} (\bibinfo {collaboration} {H1, ZEUS}),\ }\href {https://doi.org/10.1140/epjc/s10052-015-3710-4} {\bibfield  {journal} {\bibinfo  {journal} {Eur. Phys. J. C}\ }\textbf {\bibinfo {volume} {75}},\ \bibinfo {pages} {580} (\bibinfo {year} {2015})},\ \Eprint {https://arxiv.org/abs/1506.06042} {arXiv:1506.06042 [hep-ex]} \BibitemShut {NoStop}%
\bibitem [{\citenamefont {Cooper-Sarkar}\ \emph {et~al.}(2011)\citenamefont {Cooper-Sarkar}, \citenamefont {Mertsch},\ and\ \citenamefont {Sarkar}}]{Cooper-Sarkar:2011jtt}%
  \BibitemOpen
  \bibfield  {author} {\bibinfo {author} {\bibfnamefont {A.}~\bibnamefont {Cooper-Sarkar}}, \bibinfo {author} {\bibfnamefont {P.}~\bibnamefont {Mertsch}},\ and\ \bibinfo {author} {\bibfnamefont {S.}~\bibnamefont {Sarkar}},\ }\href {https://doi.org/10.1007/JHEP08(2011)042} {\bibfield  {journal} {\bibinfo  {journal} {JHEP}\ }\textbf {\bibinfo {volume} {08}},\ \bibinfo {pages} {042}},\ \Eprint {https://arxiv.org/abs/1106.3723} {arXiv:1106.3723 [hep-ph]} \BibitemShut {NoStop}%
\bibitem [{\citenamefont {Abbasi}\ \emph {et~al.}(2024{\natexlab{a}})\citenamefont {Abbasi} \emph {et~al.}}]{IceCubeCollaboration:2024dxk}%
  \BibitemOpen
  \bibfield  {author} {\bibinfo {author} {\bibfnamefont {R.}~\bibnamefont {Abbasi}} \emph {et~al.} (\bibinfo {collaboration} {IceCube}),\ }\href {https://doi.org/10.1103/PhysRevD.110.092009} {\bibfield  {journal} {\bibinfo  {journal} {Phys. Rev. D}\ }\textbf {\bibinfo {volume} {110}},\ \bibinfo {pages} {092009} (\bibinfo {year} {2024}{\natexlab{a}})},\ \Eprint {https://arxiv.org/abs/2405.08077} {arXiv:2405.08077 [hep-ex]} \BibitemShut {NoStop}%
\bibitem [{\citenamefont {Abbasi}\ \emph {et~al.}(2024{\natexlab{b}})\citenamefont {Abbasi} \emph {et~al.}}]{IceCubeCollaboration:2024nle}%
  \BibitemOpen
  \bibfield  {author} {\bibinfo {author} {\bibfnamefont {R.}~\bibnamefont {Abbasi}} \emph {et~al.} (\bibinfo {collaboration} {IceCube}),\ }\href {https://doi.org/10.1103/PhysRevLett.133.201804} {\bibfield  {journal} {\bibinfo  {journal} {Phys. Rev. Lett.}\ }\textbf {\bibinfo {volume} {133}},\ \bibinfo {pages} {201804} (\bibinfo {year} {2024}{\natexlab{b}})},\ \Eprint {https://arxiv.org/abs/2405.08070} {arXiv:2405.08070 [hep-ex]} \BibitemShut {NoStop}%
\bibitem [{\citenamefont {Goodman}\ and\ \citenamefont {Weare}(2010)}]{Goodman:2010dyf}%
  \BibitemOpen
  \bibfield  {author} {\bibinfo {author} {\bibfnamefont {J.}~\bibnamefont {Goodman}}\ and\ \bibinfo {author} {\bibfnamefont {J.}~\bibnamefont {Weare}},\ }\href {https://doi.org/10.2140/camcos.2010.5.65} {\bibfield  {journal} {\bibinfo  {journal} {Commun. Appl. Math. Comput. Sc.}\ }\textbf {\bibinfo {volume} {5}},\ \bibinfo {pages} {65} (\bibinfo {year} {2010})}\BibitemShut {NoStop}%
\bibitem [{\citenamefont {{Foreman-Mackey}}\ \emph {et~al.}(2013)\citenamefont {{Foreman-Mackey}}, \citenamefont {{Hogg}}, \citenamefont {{Lang}},\ and\ \citenamefont {{Goodman}}}]{emcee2013}%
  \BibitemOpen
  \bibfield  {author} {\bibinfo {author} {\bibfnamefont {D.}~\bibnamefont {{Foreman-Mackey}}}, \bibinfo {author} {\bibfnamefont {D.~W.}\ \bibnamefont {{Hogg}}}, \bibinfo {author} {\bibfnamefont {D.}~\bibnamefont {{Lang}}},\ and\ \bibinfo {author} {\bibfnamefont {J.}~\bibnamefont {{Goodman}}},\ }\href {https://doi.org/10.1086/670067} {\bibfield  {journal} {\bibinfo  {journal} {PASP}\ }\textbf {\bibinfo {volume} {125}},\ \bibinfo {pages} {306} (\bibinfo {year} {2013})},\ \Eprint {https://arxiv.org/abs/1202.3665} {arXiv:1202.3665 [astro-ph.IM]} \BibitemShut {NoStop}%
\bibitem [{\citenamefont {{Luzum}}\ \emph {et~al.}(2011)\citenamefont {{Luzum}}, \citenamefont {{Capitaine}}, \citenamefont {{Fienga}}, \citenamefont {{Folkner}}, \citenamefont {{Fukushima}}, \citenamefont {{Hilton}}, \citenamefont {{Hohenkerk}}, \citenamefont {{Krasinsky}}, \citenamefont {{Petit}}, \citenamefont {{Pitjeva}}, \citenamefont {{Soffel}},\ and\ \citenamefont {{Wallace}}}]{2011CeMDA.110..293L}%
  \BibitemOpen
  \bibfield  {author} {\bibinfo {author} {\bibfnamefont {B.}~\bibnamefont {{Luzum}}}, \bibinfo {author} {\bibfnamefont {N.}~\bibnamefont {{Capitaine}}}, \bibinfo {author} {\bibfnamefont {A.}~\bibnamefont {{Fienga}}}, \bibinfo {author} {\bibfnamefont {W.}~\bibnamefont {{Folkner}}}, \bibinfo {author} {\bibfnamefont {T.}~\bibnamefont {{Fukushima}}}, \bibinfo {author} {\bibfnamefont {J.}~\bibnamefont {{Hilton}}}, \bibinfo {author} {\bibfnamefont {C.}~\bibnamefont {{Hohenkerk}}}, \bibinfo {author} {\bibfnamefont {G.}~\bibnamefont {{Krasinsky}}}, \bibinfo {author} {\bibfnamefont {G.}~\bibnamefont {{Petit}}}, \bibinfo {author} {\bibfnamefont {E.}~\bibnamefont {{Pitjeva}}}, \bibinfo {author} {\bibfnamefont {M.}~\bibnamefont {{Soffel}}},\ and\ \bibinfo {author} {\bibfnamefont {P.}~\bibnamefont {{Wallace}}},\ }\href {https://doi.org/10.1007/s10569-011-9352-4} {\bibfield  {journal} {\bibinfo  {journal} {Celestial Mechanics and Dynamical Astronomy}\ }\textbf {\bibinfo {volume} {110}},\ \bibinfo {pages} {293} (\bibinfo
  {year} {2011})}\BibitemShut {NoStop}%
\bibitem [{\citenamefont {Ishihara}(2021)}]{Ishihara:2019aao}%
  \BibitemOpen
  \bibfield  {author} {\bibinfo {author} {\bibfnamefont {A.}~\bibnamefont {Ishihara}} (\bibinfo {collaboration} {IceCube}),\ }\href {https://doi.org/10.22323/1.358.1031} {\bibfield  {journal} {\bibinfo  {journal} {PoS}\ }\textbf {\bibinfo {volume} {ICRC2019}},\ \bibinfo {pages} {1031} (\bibinfo {year} {2021})},\ \Eprint {https://arxiv.org/abs/1908.09441} {arXiv:1908.09441 [astro-ph.HE]} \BibitemShut {NoStop}%
\bibitem [{\citenamefont {Aartsen}\ \emph {et~al.}(2021)\citenamefont {Aartsen} \emph {et~al.}}]{IceCube-Gen2:2020qha}%
  \BibitemOpen
  \bibfield  {author} {\bibinfo {author} {\bibfnamefont {M.~G.}\ \bibnamefont {Aartsen}} \emph {et~al.} (\bibinfo {collaboration} {IceCube-Gen2}),\ }\href {https://doi.org/10.1088/1361-6471/abbd48} {\bibfield  {journal} {\bibinfo  {journal} {J. Phys. G}\ }\textbf {\bibinfo {volume} {48}},\ \bibinfo {pages} {060501} (\bibinfo {year} {2021})},\ \Eprint {https://arxiv.org/abs/2008.04323} {arXiv:2008.04323 [astro-ph.HE]} \BibitemShut {NoStop}%
\bibitem [{\citenamefont {Adrian-Martinez}\ \emph {et~al.}(2016)\citenamefont {Adrian-Martinez} \emph {et~al.}}]{KM3Net:2016zxf}%
  \BibitemOpen
  \bibfield  {author} {\bibinfo {author} {\bibfnamefont {S.}~\bibnamefont {Adrian-Martinez}} \emph {et~al.} (\bibinfo {collaboration} {KM3NeT}),\ }\href {https://doi.org/10.1088/0954-3899/43/8/084001} {\bibfield  {journal} {\bibinfo  {journal} {J. Phys. G}\ }\textbf {\bibinfo {volume} {43}},\ \bibinfo {pages} {084001} (\bibinfo {year} {2016})},\ \Eprint {https://arxiv.org/abs/1601.07459} {arXiv:1601.07459 [astro-ph.IM]} \BibitemShut {NoStop}%
\bibitem [{\citenamefont {Halzen}\ and\ \citenamefont {Saltzberg}(1998)}]{Halzen:1998be}%
  \BibitemOpen
  \bibfield  {author} {\bibinfo {author} {\bibfnamefont {F.}~\bibnamefont {Halzen}}\ and\ \bibinfo {author} {\bibfnamefont {D.}~\bibnamefont {Saltzberg}},\ }\href {https://doi.org/10.1103/PhysRevLett.81.4305} {\bibfield  {journal} {\bibinfo  {journal} {Phys. Rev. Lett.}\ }\textbf {\bibinfo {volume} {81}},\ \bibinfo {pages} {4305} (\bibinfo {year} {1998})},\ \Eprint {https://arxiv.org/abs/hep-ph/9804354} {arXiv:hep-ph/9804354} \BibitemShut {NoStop}%
\bibitem [{\citenamefont {Ya{\~n}ez}\ and\ \citenamefont {Fedynitch}(2023)}]{Yanez:2023lsy}%
  \BibitemOpen
  \bibfield  {author} {\bibinfo {author} {\bibfnamefont {J.~P.}\ \bibnamefont {Ya{\~n}ez}}\ and\ \bibinfo {author} {\bibfnamefont {A.}~\bibnamefont {Fedynitch}},\ }\href {https://doi.org/10.1103/PhysRevD.107.123037} {\bibfield  {journal} {\bibinfo  {journal} {Phys. Rev. D}\ }\textbf {\bibinfo {volume} {107}},\ \bibinfo {pages} {123037} (\bibinfo {year} {2023})},\ \Eprint {https://arxiv.org/abs/2303.00022} {arXiv:2303.00022 [hep-ph]} \BibitemShut {NoStop}%
\bibitem [{\citenamefont {{Picone}}\ \emph {et~al.}(2002)\citenamefont {{Picone}}, \citenamefont {{Hedin}}, \citenamefont {{Drob}},\ and\ \citenamefont {{Aikin}}}]{2002JGRA..107.1468P}%
  \BibitemOpen
  \bibfield  {author} {\bibinfo {author} {\bibfnamefont {J.~M.}\ \bibnamefont {{Picone}}}, \bibinfo {author} {\bibfnamefont {A.~E.}\ \bibnamefont {{Hedin}}}, \bibinfo {author} {\bibfnamefont {D.~P.}\ \bibnamefont {{Drob}}},\ and\ \bibinfo {author} {\bibfnamefont {A.~C.}\ \bibnamefont {{Aikin}}},\ }\href {https://doi.org/10.1029/2002JA009430} {\bibfield  {journal} {\bibinfo  {journal} {Journal of Geophysical Research (Space Physics)}\ }\textbf {\bibinfo {volume} {107}},\ \bibinfo {eid} {1468} (\bibinfo {year} {2002})}\BibitemShut {NoStop}%
\bibitem [{\citenamefont {Riehn}\ \emph {et~al.}(2020)\citenamefont {Riehn}, \citenamefont {Engel}, \citenamefont {Fedynitch}, \citenamefont {Gaisser},\ and\ \citenamefont {Stanev}}]{Riehn:2019jet}%
  \BibitemOpen
  \bibfield  {author} {\bibinfo {author} {\bibfnamefont {F.}~\bibnamefont {Riehn}}, \bibinfo {author} {\bibfnamefont {R.}~\bibnamefont {Engel}}, \bibinfo {author} {\bibfnamefont {A.}~\bibnamefont {Fedynitch}}, \bibinfo {author} {\bibfnamefont {T.~K.}\ \bibnamefont {Gaisser}},\ and\ \bibinfo {author} {\bibfnamefont {T.}~\bibnamefont {Stanev}},\ }\href {https://doi.org/10.1103/PhysRevD.102.063002} {\bibfield  {journal} {\bibinfo  {journal} {Phys. Rev. D}\ }\textbf {\bibinfo {volume} {102}},\ \bibinfo {pages} {063002} (\bibinfo {year} {2020})},\ \Eprint {https://arxiv.org/abs/1912.03300} {arXiv:1912.03300 [hep-ph]} \BibitemShut {NoStop}%
\bibitem [{\citenamefont {Klein}\ \emph {et~al.}(2020)\citenamefont {Klein}, \citenamefont {Robertson},\ and\ \citenamefont {Vogt}}]{Klein:2020nuk}%
  \BibitemOpen
  \bibfield  {author} {\bibinfo {author} {\bibfnamefont {S.~R.}\ \bibnamefont {Klein}}, \bibinfo {author} {\bibfnamefont {S.~A.}\ \bibnamefont {Robertson}},\ and\ \bibinfo {author} {\bibfnamefont {R.}~\bibnamefont {Vogt}},\ }\href {https://doi.org/10.1103/PhysRevC.102.015808} {\bibfield  {journal} {\bibinfo  {journal} {Phys. Rev. C}\ }\textbf {\bibinfo {volume} {102}},\ \bibinfo {pages} {015808} (\bibinfo {year} {2020})},\ \Eprint {https://arxiv.org/abs/2001.03677} {arXiv:2001.03677 [hep-ph]} \BibitemShut {NoStop}%
\bibitem [{\citenamefont {Weigel}\ \emph {et~al.}(2025)\citenamefont {Weigel}, \citenamefont {Conrad},\ and\ \citenamefont {Garcia-Soto}}]{Weigel:2024gzh}%
  \BibitemOpen
  \bibfield  {author} {\bibinfo {author} {\bibfnamefont {P.~L.~R.}\ \bibnamefont {Weigel}}, \bibinfo {author} {\bibfnamefont {J.~M.}\ \bibnamefont {Conrad}},\ and\ \bibinfo {author} {\bibfnamefont {A.}~\bibnamefont {Garcia-Soto}},\ }\href {https://doi.org/10.1103/PhysRevD.111.043044} {\bibfield  {journal} {\bibinfo  {journal} {Phys. Rev. D}\ }\textbf {\bibinfo {volume} {111}},\ \bibinfo {pages} {043044} (\bibinfo {year} {2025})},\ \Eprint {https://arxiv.org/abs/2408.05866} {arXiv:2408.05866 [hep-ph]} \BibitemShut {NoStop}%
\bibitem [{\citenamefont {Eskola}\ \emph {et~al.}(2022)\citenamefont {Eskola}, \citenamefont {Paakkinen}, \citenamefont {Paukkunen},\ and\ \citenamefont {Salgado}}]{Eskola:2021nhw}%
  \BibitemOpen
  \bibfield  {author} {\bibinfo {author} {\bibfnamefont {K.~J.}\ \bibnamefont {Eskola}}, \bibinfo {author} {\bibfnamefont {P.}~\bibnamefont {Paakkinen}}, \bibinfo {author} {\bibfnamefont {H.}~\bibnamefont {Paukkunen}},\ and\ \bibinfo {author} {\bibfnamefont {C.~A.}\ \bibnamefont {Salgado}},\ }\href {https://doi.org/10.1140/epjc/s10052-022-10359-0} {\bibfield  {journal} {\bibinfo  {journal} {Eur. Phys. J. C}\ }\textbf {\bibinfo {volume} {82}},\ \bibinfo {pages} {413} (\bibinfo {year} {2022})},\ \Eprint {https://arxiv.org/abs/2112.12462} {arXiv:2112.12462 [hep-ph]} \BibitemShut {NoStop}%
\bibitem [{\citenamefont {Aartsen}\ \emph {et~al.}(2019)\citenamefont {Aartsen} \emph {et~al.}}]{IceCube:2019lxi}%
  \BibitemOpen
  \bibfield  {author} {\bibinfo {author} {\bibfnamefont {M.~G.}\ \bibnamefont {Aartsen}} \emph {et~al.} (\bibinfo {collaboration} {IceCube}),\ }\href {https://doi.org/10.1088/1475-7516/2019/10/048} {\bibfield  {journal} {\bibinfo  {journal} {JCAP}\ }\textbf {\bibinfo {volume} {10}},\ \bibinfo {pages} {048}},\ \Eprint {https://arxiv.org/abs/1909.01530} {arXiv:1909.01530 [hep-ex]} \BibitemShut {NoStop}%
\bibitem [{\citenamefont {Abbasi}\ \emph {et~al.}(2025)\citenamefont {Abbasi} \emph {et~al.}}]{IceCube:2025yvq}%
  \BibitemOpen
  \bibfield  {author} {\bibinfo {author} {\bibfnamefont {R.}~\bibnamefont {Abbasi}} \emph {et~al.} (\bibinfo {collaboration} {IceCube}),\ }\href@noop {} {\  (\bibinfo {year} {2025})},\ \Eprint {https://arxiv.org/abs/2506.04491} {arXiv:2506.04491 [hep-ex]} \BibitemShut {NoStop}%
\bibitem [{\citenamefont {Gelman}\ \emph {et~al.}(1996)\citenamefont {Gelman}, \citenamefont {Meng},\ and\ \citenamefont {Stern}}]{Gelman:1996}%
  \BibitemOpen
  \bibfield  {author} {\bibinfo {author} {\bibfnamefont {A.}~\bibnamefont {Gelman}}, \bibinfo {author} {\bibfnamefont {X.-L.}\ \bibnamefont {Meng}},\ and\ \bibinfo {author} {\bibfnamefont {H.}~\bibnamefont {Stern}}\ }(\bibinfo  {publisher} {Institute of Statistical Science, Academia Sinica},\ \bibinfo {year} {1996})\ pp.\ \bibinfo {pages} {733--760}\BibitemShut {NoStop}%
\bibitem [{\citenamefont {Acero}\ \emph {et~al.}(2024)\citenamefont {Acero} \emph {et~al.}}]{NOvA:2023iam}%
  \BibitemOpen
  \bibfield  {author} {\bibinfo {author} {\bibfnamefont {M.~A.}\ \bibnamefont {Acero}} \emph {et~al.} (\bibinfo {collaboration} {NOvA}),\ }\href {https://doi.org/10.1103/PhysRevD.110.012005} {\bibfield  {journal} {\bibinfo  {journal} {Phys. Rev. D}\ }\textbf {\bibinfo {volume} {110}},\ \bibinfo {pages} {012005} (\bibinfo {year} {2024})},\ \Eprint {https://arxiv.org/abs/2311.07835} {arXiv:2311.07835 [hep-ex]} \BibitemShut {NoStop}%
\bibitem [{\citenamefont {Aartsen}\ \emph {et~al.}(2020)\citenamefont {Aartsen} \emph {et~al.}}]{IceCube:2020acn}%
  \BibitemOpen
  \bibfield  {author} {\bibinfo {author} {\bibfnamefont {M.~G.}\ \bibnamefont {Aartsen}} \emph {et~al.} (\bibinfo {collaboration} {IceCube}),\ }\href {https://doi.org/10.1103/PhysRevLett.125.121104} {\bibfield  {journal} {\bibinfo  {journal} {Phys. Rev. Lett.}\ }\textbf {\bibinfo {volume} {125}},\ \bibinfo {pages} {121104} (\bibinfo {year} {2020})},\ \Eprint {https://arxiv.org/abs/2001.09520} {arXiv:2001.09520 [astro-ph.HE]} \BibitemShut {NoStop}%
\bibitem [{\citenamefont {Abbasi}\ \emph {et~al.}(2021{\natexlab{b}})\citenamefont {Abbasi} \emph {et~al.}}]{IceCube:2020wum}%
  \BibitemOpen
  \bibfield  {author} {\bibinfo {author} {\bibfnamefont {R.}~\bibnamefont {Abbasi}} \emph {et~al.} (\bibinfo {collaboration} {IceCube}),\ }\href {https://doi.org/10.1103/PhysRevD.104.022002} {\bibfield  {journal} {\bibinfo  {journal} {Phys. Rev. D}\ }\textbf {\bibinfo {volume} {104}},\ \bibinfo {pages} {022002} (\bibinfo {year} {2021}{\natexlab{b}})},\ \Eprint {https://arxiv.org/abs/2011.03545} {arXiv:2011.03545 [astro-ph.HE]} \BibitemShut {NoStop}%
\bibitem [{\citenamefont {Abbasi}\ \emph {et~al.}(2022)\citenamefont {Abbasi} \emph {et~al.}}]{IceCube:2021uhz}%
  \BibitemOpen
  \bibfield  {author} {\bibinfo {author} {\bibfnamefont {R.}~\bibnamefont {Abbasi}} \emph {et~al.} (\bibinfo {collaboration} {IceCube}),\ }\href {https://doi.org/10.3847/1538-4357/ac4d29} {\bibfield  {journal} {\bibinfo  {journal} {Astrophys. J.}\ }\textbf {\bibinfo {volume} {928}},\ \bibinfo {pages} {50} (\bibinfo {year} {2022})},\ \Eprint {https://arxiv.org/abs/2111.10299} {arXiv:2111.10299 [astro-ph.HE]} \BibitemShut {NoStop}%
\end{thebibliography}%
\clearpage

\section{Methods}

The data selection and reconstruction procedures are described in detail in Refs.~\cite{IceCubeCollaboration:2024dxk,IceCubeCollaboration:2024nle}; we summarize the key elements relevant for this analysis here. 
We analyze muon tracks produced in muon-neutrino charged-current interactions. 
These events are classified as \textit{starting} or \textit{throughgoing} depending on whether the neutrino interaction occurs within the instrumented volume of IceCube or outside it, with only the resulting muon entering the detector. 
Each event is characterized by a reconstructed muon energy, $E_\mu$, as a proxy for $E_\nu$, the neutrino energy, and a reconstructed direction given by the zenith angle $\theta$ and azimuth $\phi$. 
As shown in Fig.~\ref{fig:schematic}, $\theta$ is defined relative to the local vertical at the South Pole. 
We restrict the sample to upgoing tracks, $-1 < \cos\theta < 0$, to select for those neutrinos which have traversed the Earth.

The dataset spans 10.7 years of IceCube 86-string operation and comprises $368{,}071$ tracks in total, including $93{,}762$ reconstructed starting and $274{,}309$ reconstructed throughgoing events, collected over the time period from 13 May, 2011 to 7 June, 2022. 
The reconstructed neutrino energy range is $500\,\mathrm{GeV}$ to $100\,\mathrm{TeV}$. 
Events are binned in a two-dimensional histogram in reconstructed energy and $\cos\theta$, and a binned-likelihood analysis is performed to infer the parameters describing the Earth's radial density profile.
We split the energy range into 22 equally log-spaced bins (ranging from $500\:\textrm{GeV}$ to $100\:\textrm{TeV}$) and the $\cos\theta$ range into 50 equally linearly-spaced bins (ranging from $-1$ to $0$); the starting and throughgoing events are fit separately. 

For the density parametrization, we approximate the Earth as a set of uniform-density, concentric, spherically-symmetric shells.
This analysis uses five shells, with boundaries placed at the inner-core/outer-core and core–mantle transitions and with additional shells equally spaced in radius within the outer-core and mantle regions. 
No other external geophysical constraints are imposed or assumed; rather, the Earth's total mass and moment of inertia are derived from the posterior density samples.
The density parameters are assigned nonnegative, uniform priors in linear space.
As an extension providing finer radial granularity, we also consider an eight-shell parametrization, described in Supplementary Materials.
To model the effect of varying the Earth density profile for a given flux component, we use the \texttt{nuFATE}~\cite{Vincent:2017svp} software to construct a spline for neutrino attenuation as a function of column depth (in $\mathrm{g\,m^{-2}}$) and neutrino energy (in $\textrm{GeV}$).
To compute the expected attenuation for a given density profile, we compute the column depth for each neutrino direction and read the spline to obtain the attenuation factor.
We can then smoothly vary the density of the Earth layers and obtain a prediction for the corresponding change in the neutrino flux.
For simplicity, we do not take into account second-order flux effects such as tau regeneration~\cite{Halzen:1998be}, since these effects enter at the percent level, and are negligibly small compared to uncertainties on the inferred Earth densities.

While inferring the density parameters, we account for systematic uncertainties through a set of 37 nuisance parameters that modify the predicted neutrino distributions. 
These parameters are assigned priors and allowed ranges informed by external measurements and calibration studies. 
We summarize their structure here; a complete list of parameters and prior definitions is provided in the Supplementary Materials.

The nuisance parameters are grouped into five categories: overall normalization; conventional atmospheric neutrino flux uncertainties; non-conventional flux uncertainties; neutrino attenuation cross section uncertainties; and detector-related effects.
The conventional atmospheric flux (arising mostly from pion and kaon decay in air showers) is modelled using the \texttt{DaemonFlux} model~\cite{Yanez:2023lsy}, which includes parameters describing uncertainties in the primary cosmic-ray spectrum and in hadronic interactions within atmospheric showers, and assumes the \texttt{NRLMSISE-00} model~\cite{2002JGRA..107.1468P} to describe the atmosphere. 
A subset of these \texttt{DaemonFlux} parameters, with associated priors and ranges, is included in the fit. 
We additionally allow for variations in the atmospheric density profile and in the kaon energy-loss rate in air.
The non-conventional flux consists of an astrophysical component and a prompt atmospheric component (the latter arising from charmed meson decays in the atmosphere). 
The astrophysical flux is modeled as a broken power law characterized by a normalization, two spectral indices, and a pivot energy, while the prompt component baseline is the \texttt{DaemonFlux} output (based on the \texttt{SIBYLL 2.3d} model~\cite{Riehn:2019jet}), allowed to vary with a free normalization parameter.

Uncertainties in the neutrino attenuation are implemented as scaling factors on the adopted CSMS cross section calculation at NLO in the Standard Model~\cite{Cooper-Sarkar:2011jtt}, based on the HERA PDF \cite{H1:2009pze,H1:2015ubc}. 
Notably, while changes in the cross section are degenerate with density changes at a given zenith angle, this degeneracy can be broken given the full angular distribution.
Nuclear effects, including low-Bjorken-$x$ shadowing, can modify neutrino–nucleus cross sections through changes to the underlying parton distribution functions~\cite{Klein:2020nuk,Weigel:2024gzh}.
To see if this has an effect beyond what can be described using the cross section nuisance parameters, we implemented nuclear corrections in the simulation following Ref.~\cite{Weigel:2024gzh}, which incorporates the EPPS21 nuclear PDF set~\cite{Eskola:2021nhw}, and analyzed the effect on density measurement sensitivities.
We found that the impact on our overall density measurement sensitivities is minimal, and can be adequately described by our current treatment of neutrino attenuation cross section uncertainties.
A related correction not explicitly tested here is target non-isoscalarity — the neutron excess present in Earth nuclei relative to the equal proton–neutron mixture assumed by CSMS. 
This is expected to be of comparable magnitude to the shadowing correction, and therefore should not have a significant impact on the present results, but nonetheless should be incorporated in future higher-precision analyses.

Detector-related systematics include uncertainties in the optical properties of the glacial ice and in the response of the DOMs. 
DOM response is parameterized by an overall efficiency factor. 
Variations in the bulk ice are described using the SnowStorm method~\cite{IceCube:2019lxi}, in which Fourier modes parameterize spatial variations in photon scattering and absorption lengths, along with an additional parameter accounting for optical effects in the refrozen borehole ice surrounding the DOMs.

Based on studies with simulation, we expect that astrophysical flux parameters contribute the most to the systematic uncertainty, with all other sources being subleading. 
This interpretation is reasonable, since the astrophysical flux is poorly-constrained compared to the conventional atmospheric flux, and its large uncertainty will drive a corresponding uncertainty in the density inference. 
Overall, in the overwhelming majority of analysis bins, the systematic uncertainty is dominant, compared to the statistical uncertainty.

To perform the inference, we employ a Markov Chain Monte Carlo (MCMC) approach implemented with the \texttt{emcee} ensemble sampler~\cite{emcee2013}. 
We use a combination of differential evolution and stretch moves to efficiently explore the high-dimensional parameter space. 
The likelihood function is evaluated using the \texttt{GollumFit} framework~\cite{IceCube:2025yvq}, which incorporates the full set of nuisance parameters together with the Earth density parametrization.
We assessed the convergence of the MCMC chains by estimating the integrated autocorrelation time following Ref.~\cite{emcee2013}, ensuring that chains are evolved for at least 50 autocorrelation times to ensure adequate sampling of the posterior. 
The resulting chains provide samples of the joint posterior distribution over both density and nuisance parameters. 
We report posterior modes as central values and quote highest-posterior-density (HPD) credible intervals as uncertainties.

\begin{figure*}
    \centering
    \includegraphics[width=1.0\linewidth]{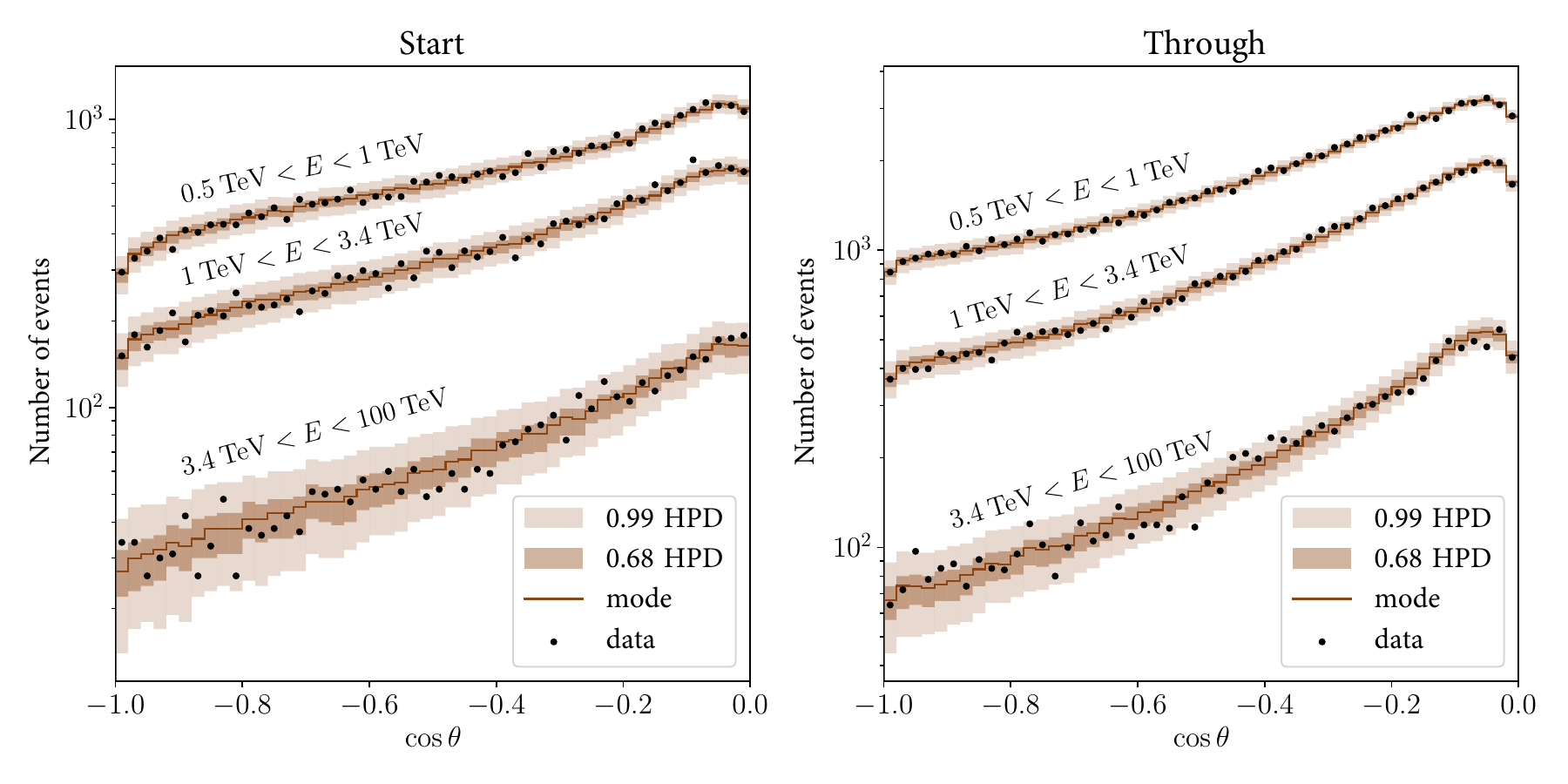}
    \caption{\textbf{Data and best-fit $\cos\theta$ distributions for starting and throughgoing tracks across three energy ranges.}
    The $\cos\theta$ distributions for data (black points) and the best-fit posterior (brown bands) are shown for $0.5<E_\nu<1\,\mathrm{TeV}$, $1<E_\nu<3.4\,\mathrm{TeV}$, and $3.4<E_\nu<100\,\mathrm{TeV}$. 
    The brown bands represent posterior predictive distributions, obtained by evaluating the simulated distribution at each posterior sample (over density and nuisance parameters), applying Poisson fluctuations, and aggregating the results. 
    The solid brown line indicates the posterior mode, while the dark and light brown bands show the 68\% and 99\% HPD intervals, respectively.
    }
    \label{fig:split_dist}
\end{figure*}

Goodness-of-fit is evaluated using a posterior predictive $p$-value approach introduced in Ref.~\cite{Gelman:1996} and implemented in, for instance, Ref.~\cite{NOvA:2023iam}. 
For the five-shell analysis including all data, we obtain $p=16\%$, indicating good agreement between the model and the observed data. 
Fig.~\ref{fig:split_dist} shows representative comparisons of the reconstructed $\cos\theta$ distributions (which are one-dimensional projections of the two-dimensional histograms in reconstructed energy and zenith) in three energy intervals, where the data (black points) are overlaid with posterior predictive distributions (brown bands showing the mode, 68\% HPD, and 99\% HPD regions). 
Across all energy ranges, the data are consistent with the predictive distributions within the combined statistical and systematic uncertainties.

\section{Acknowledgements}
The authors gratefully acknowledge the support from the following agencies and institutions:
USA {\textendash} U.S. National Science Foundation-Office of Polar Programs,
U.S. National Science Foundation-Physics Division,
U.S. National Science Foundation-EPSCoR,
U.S. National Science Foundation-Office of Advanced Cyberinfrastructure,
Wisconsin Alumni Research Foundation,
Center for High Throughput Computing (CHTC) at the University of Wisconsin{\textendash}Madison,
Open Science Grid (OSG),
Partnership to Advance Throughput Computing (PATh),
Advanced Cyberinfrastructure Coordination Ecosystem: Services {\&} Support (ACCESS),
Frontera and Ranch computing project at the Texas Advanced Computing Center,
U.S. Department of Energy-National Energy Research Scientific Computing Center,
Particle astrophysics research computing center at the University of Maryland,
Michigan State University,
Astroparticle physics computational facility at Marquette University,
NVIDIA Corporation,
and Google Cloud Platform;
Belgium {\textendash} Funds for Scientific Research (FRS-FNRS and FWO),
FWO Odysseus and Big Science programmes,
and Belgian Federal Science Policy Office (Belspo);
Germany {\textendash} Bundesministerium f{\"u}r Forschung, Technologie und Raumfahrt (BMFTR),
Deutsche Forschungsgemeinschaft (DFG),
Helmholtz Alliance for Astroparticle Physics (HAP),
Initiative and Networking Fund of the Helmholtz Association,
Deutsches Elektronen Synchrotron (DESY),
and High Performance Computing cluster of the RWTH Aachen;
Sweden {\textendash} Swedish Research Council,
Swedish Polar Research Secretariat,
Swedish National Infrastructure for Computing (SNIC),
and Knut and Alice Wallenberg Foundation;
European Union {\textendash} EGI Advanced Computing for research;
Australia {\textendash} Australian Research Council;
Canada {\textendash} Natural Sciences and Engineering Research Council of Canada,
Calcul Qu{\'e}bec, Compute Ontario, Canada Foundation for Innovation, WestGrid, and Digital Research Alliance of Canada;
Denmark {\textendash} Villum Fonden, Carlsberg Foundation, and European Commission;
New Zealand {\textendash} Marsden Fund;
Japan {\textendash} Japan Society for Promotion of Science (JSPS), Ministry of Education, Culture, Sports, Science and Technology (MEXT), and Institute for Global Prominent Research (IGPR) of Chiba University;
Korea {\textendash} National Research Foundation of Korea (NRF);
Switzerland {\textendash} Swiss National Science Foundation (SNSF).

\section{Author Contributions}

The IceCube neutrino observatory was constructed and is maintained by the IceCube Collaboration. 
A large number of authors contributed to the data processing, detector calibration and Monte Carlo simulations used in this work. 
The IceCube collaboration acknowledges the substantial contributions to this manuscript from the Harvard University, Massachusetts
Institute of Technology, and University of Delaware groups. 
The final manuscript was reviewed and approved by all authors.

\section{Competing Interests}

The authors declare no competing interests.

\clearpage

\section{Supplementary Materials}

\subsection{Nuisance parameters}

Table~\ref{tab:systematics} gives the full list of nuisance parameters included in the analysis, along with their prior centers, widths, allowed ranges, and prior shapes. 
As described in the Methods, these nuisance parameters enter the binned-likelihood model by modifying the predicted energy- and zenith-dependent neutrino distributions, and are simultaneously inferred with the Earth density parameters in the joint Bayesian fit. 
The parameters are grouped into five broad categories: an overall event-rate normalization; detector response and ice optical uncertainties; conventional atmospheric flux uncertainties; non-conventional (high-energy) flux uncertainties; and neutrino--nucleon cross section uncertainties.
Unless otherwise noted, nuisance parameters are assigned Gaussian priors centered on nominal values with widths and truncation ranges motivated by external constraints and calibration studies.
Most of the \texttt{DaemonFlux}~\cite{Yanez:2023lsy} nuisance parameters (all of the parameters for the Conventional Atmospheric Flux except $\rho_\mathrm{atm}$ and $\sigma_\mathrm{K-Air}$) and all of the Ice Amplitude and Ice Phase parameters come with correlation priors, which are also included in the fit.
For a more complete discussion, refer to Ref.~\cite{IceCubeCollaboration:2024dxk}. 

The overall normalization parameter, $\mathrm{Norm}$, captures residual uncertainties in the total event rate that are not fully accounted for by other components. 
This includes contributions from flux normalization, detector acceptance, neutrino interaction cross sections at the detector, and muon propagation effects. 
In practice, it acts as a global scaling of the predicted event counts across all bins. 

\begin{figure*}
    \centering
    \includegraphics[width=1.0\linewidth]{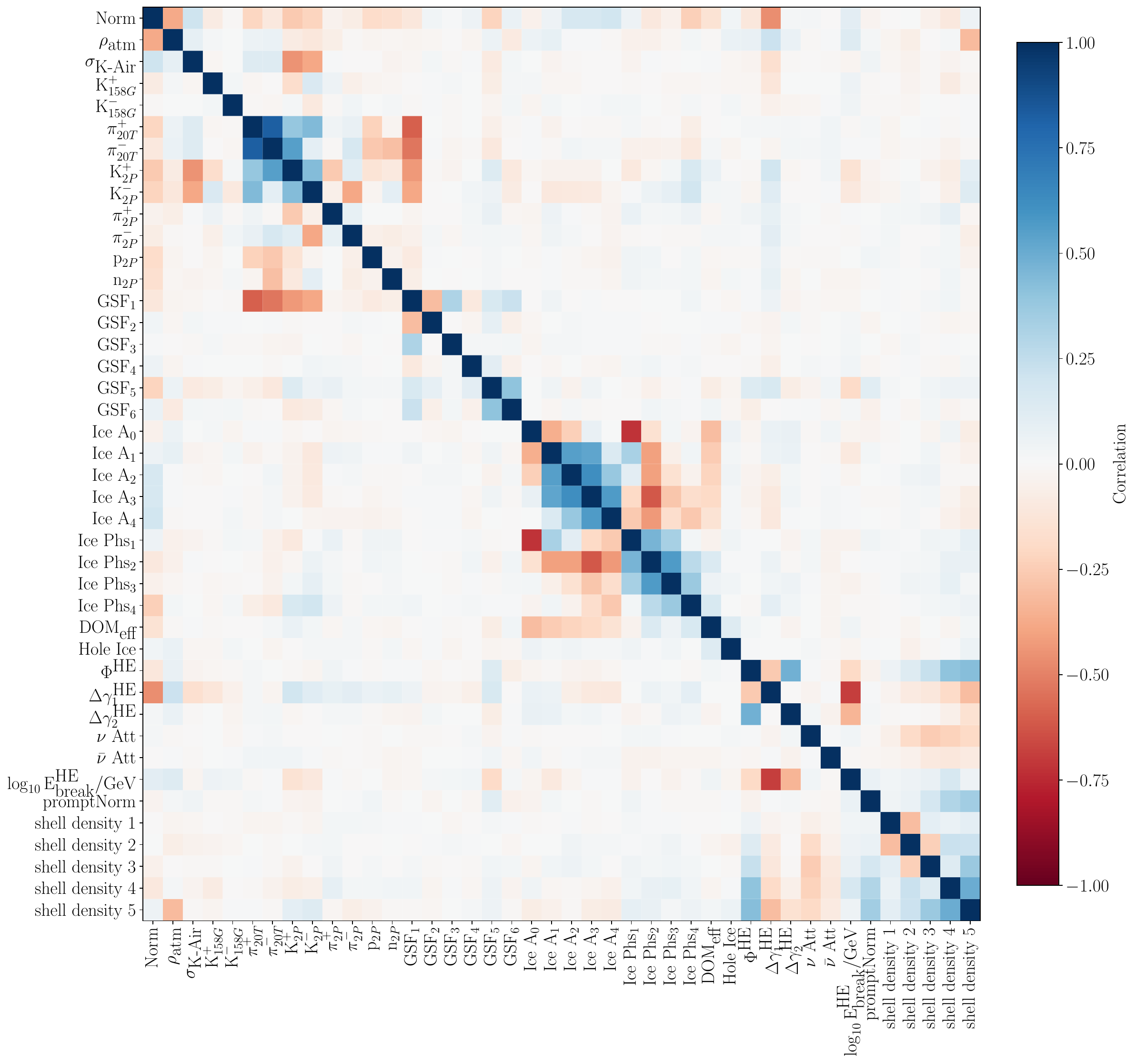}
    \caption{\textbf{The correlation matrix showing the approximate relationships in the joint posterior between all nuisance parameters and density parameters in the five-shell fit.} 
    The nuisance parameters shown here are listed in Table~\ref{tab:systematics}.
    Prior correlations within the \texttt{DaemonFlux} hadronic interaction parameters,  \texttt{DaemonFlux} cosmic ray spectrum parameters, and Ice Amplitude and Ice Phase parameters, which are incorporated in the fit, are also observed in the posteriors here.
    Remaining correlations, such as in the astrophysical flux parameters and within the density parameters, are first-order approximations of relationships that are seen in the data.
    }
    \label{fig:correlation_matrix}
\end{figure*}

Detector-related uncertainties primarily affect the mapping between true and reconstructed observables through modifications to light propagation and detection efficiency, informed by calibration data measurements.
These include uncertainties in the optical properties of the bulk ice, the local refrozen “hole ice” surrounding the optical modules (described by the parameter Hole Ice), and the overall photon detection efficiency of the sensors (described by the parameter $\mathrm{DOM_{eff}}$). 
The bulk ice uncertainties are parameterized using a reduced set~\cite{IceCube:2019lxi} of Fourier modes describing variations in scattering and absorption as a function of depth, with correlated priors reflecting their common physical origin and calibration data.
These modes are described by the parameters $\mathrm{Ice~A_{0-4}}$ and $\mathrm{Ice~Phs_{1-4}}$, which represent the amplitudes and phases associated with the decomposition of scattering and absorption variations, and variations on the simulation are encoded efficiently using the method outlined in Ref.~\cite{IceCube:2019lxi}.
These effects can differ between starting and throughgoing events due to differences in light containment and event morphology, and therefore play a role in shaping the observed distributions.
In the fit, the Hole Ice posterior accumulates near the upper boundary of its allowed range (see Table~\ref{tab:systematics}).
The correlation matrix (Fig.~\ref{fig:correlation_matrix}) shows only limited coupling between Hole Ice and the inferred density parameters.
To verify robustness, we repeated the fit with modified Hole Ice values, and found no significant change in the inferred density results.

The conventional atmospheric neutrino flux, arising from pion and kaon decays in cosmic-ray air showers, constitutes the dominant component of the event sample. 
We model its uncertainty through a set of parameters describing hadronic production yields, the primary cosmic-ray spectrum, and atmospheric properties. 
In particular, hadronic production uncertainties (described by the parameters $\mathrm{K}_{158G}^{+}$, $\mathrm{K}_{158G}^{-}$, $\pi_{20T}^{+}$, $\pi_{20T}^{-}$, $\mathrm{K}_{2P}^{+}$, $\mathrm{K}_{2P}^{-}$, $\pi_{2P}^{+}$, $\pi_{2P}^{-}$, $\mathrm{p}_{2P}$, $\mathrm{n}_{2P}$)
and cosmic-ray spectrum uncertainties (described by the parameters $\mathrm{GSF_{1-6}}$) are implemented by the flux model we use, \texttt{DaemonFlux}~\cite{Yanez:2023lsy}, and have correlated priors, reflecting shared underlying physics inputs and experimental constraints. 
The atmospheric properties ($\rho_\mathrm{atm}$ and $\sigma_\mathrm{K-Air}$) are parameters that describe uncertainties associated with a varying atmospheric temperature and the rate of kaon energy loss in air.

The non-conventional flux component includes both prompt atmospheric neutrinos and astrophysical neutrinos, which become increasingly relevant at higher energies. 
Given the modeling uncertainties in this regime, a flexible parameterization is adopted, typically allowing variations in normalization and spectral shape through a broken power-law description.
To this end, $\Phi^\mathrm{HE}$ describes a normalization, $\Delta\gamma_1^\mathrm{HE}$ and $\Delta\gamma_2^\mathrm{HE}$ describe variations in spectral indices, and $\log_{10}E_\mathrm{break}^\mathrm{HE}/\mathrm{GeV}$ describes the break point of the broken power law. 
The priors for these parameters are based on other IceCube measurements of the astrophysical flux using samples~\cite{IceCube:2020acn, IceCube:2020wum, IceCube:2021uhz} which contain significant numbers of orthogonal events, meaning events that do not overlap with the selection of tracks used in this work (for instance, cascades).
Meanwhile, the prompt atmospheric flux is allowed to vary with a normalization described by $\mathrm{promptNorm}$.
As a check, we repeated the fit with priors on the astrophysical flux normalization and spectral index parameters broadened by a factor of 10.
The inferred shell densities remain consistent with our standard fit, indicating that the result is robust to the assumed astrophysical flux.

Uncertainties in neutrino–nucleon cross sections enter via energy- and zenith-dependent attenuation as neutrinos propagate through the Earth. 
Additional attenuation nuisance parameters, $\nu \: \mathrm{Att}$ and $\bar\nu \: \mathrm{Att}$, are introduced to describe uncertainties in the attenuation of neutrinos and antineutrinos separately, particularly at higher energies, where cross section effects like nuclear shadowing are more relevant.
These parameters modify the zenith-dependent suppression of the flux and are therefore partially degenerate with an overall scaling of Earth density. 
However, this degeneracy is broken given that density parameters are permitted to vary independently of each other in the inference.

\begin{figure*}
    \centering
    \includegraphics[width=1.0\linewidth]{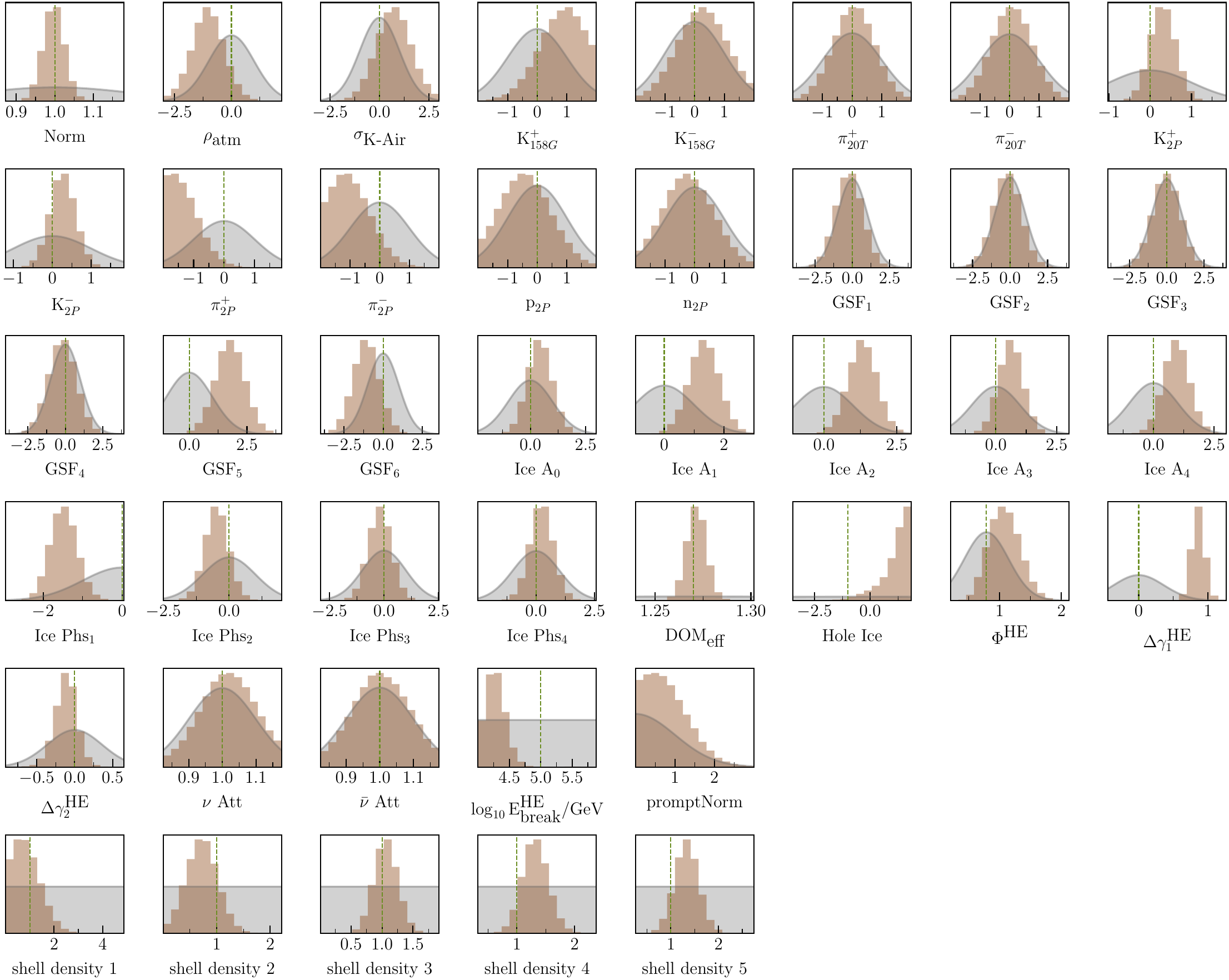}
    \caption{\textbf{The posterior distributions of nuisance parameters and density parameters, compared to their priors.} The posterior distributions are in brown while the central prior value is shown in the vertical dashed line. The prior shape is shown in grey. Nuisance parameters that have a large overlap between the posterior and the prior are strongly informed by the prior, while the parameters with very different prior and posterior distributions are strongly informed by the data. In particular, the Earth shell densities have uniform priors and Gaussian-like posteriors.}
    \label{fig:posterior_grid}
\end{figure*}

Figure~\ref{fig:posterior_grid} shows the marginalized posterior distribution for each nuisance parameter in Table~\ref{tab:systematics} together with the five Earth density parameters for the five-shell analysis. 
In this figure the posterior distributions are shown in brown, while the prior shapes are shown in grey; the prior central values are indicated by vertical dashed lines. 
Parameters for which the posterior closely follows the prior are primarily prior-constrained, whereas parameters that exhibit substantial shifts or narrowing relative to the prior are informed by the IceCube data through the fit. 
The shell densities are assigned uniform priors over their allowed physical ranges and yield approximately Gaussian-like posteriors, reflecting the constraining power of the energy- and zenith-dependent attenuation pattern.

\begin{figure*}
    \centering
    \includegraphics[width=1.0\linewidth]{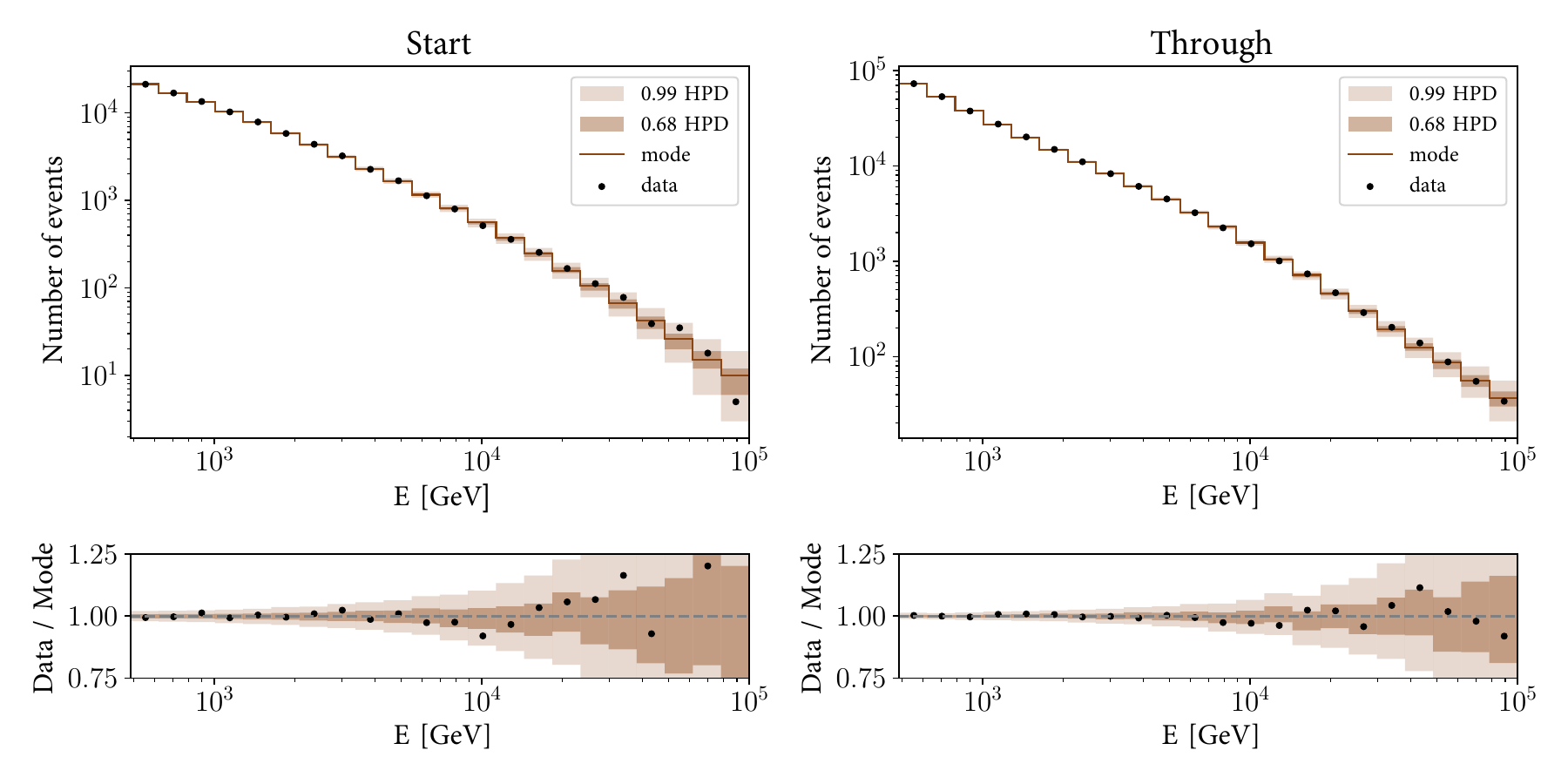}
    \caption{\textbf{The energy distribution, comparing data and best fit posteriors.} In the top panels, the data (black points) are overlaid with posterior predictive distributions (brown bands showing the mode, 68\% HPD, and 99\% HPD regions). The bottom panels show the ratio relative to the mode of the posterior distribution.}
    \label{fig:energy_dist}
\end{figure*}

\begin{figure*}
    \centering
    \includegraphics[width=1.0\linewidth]{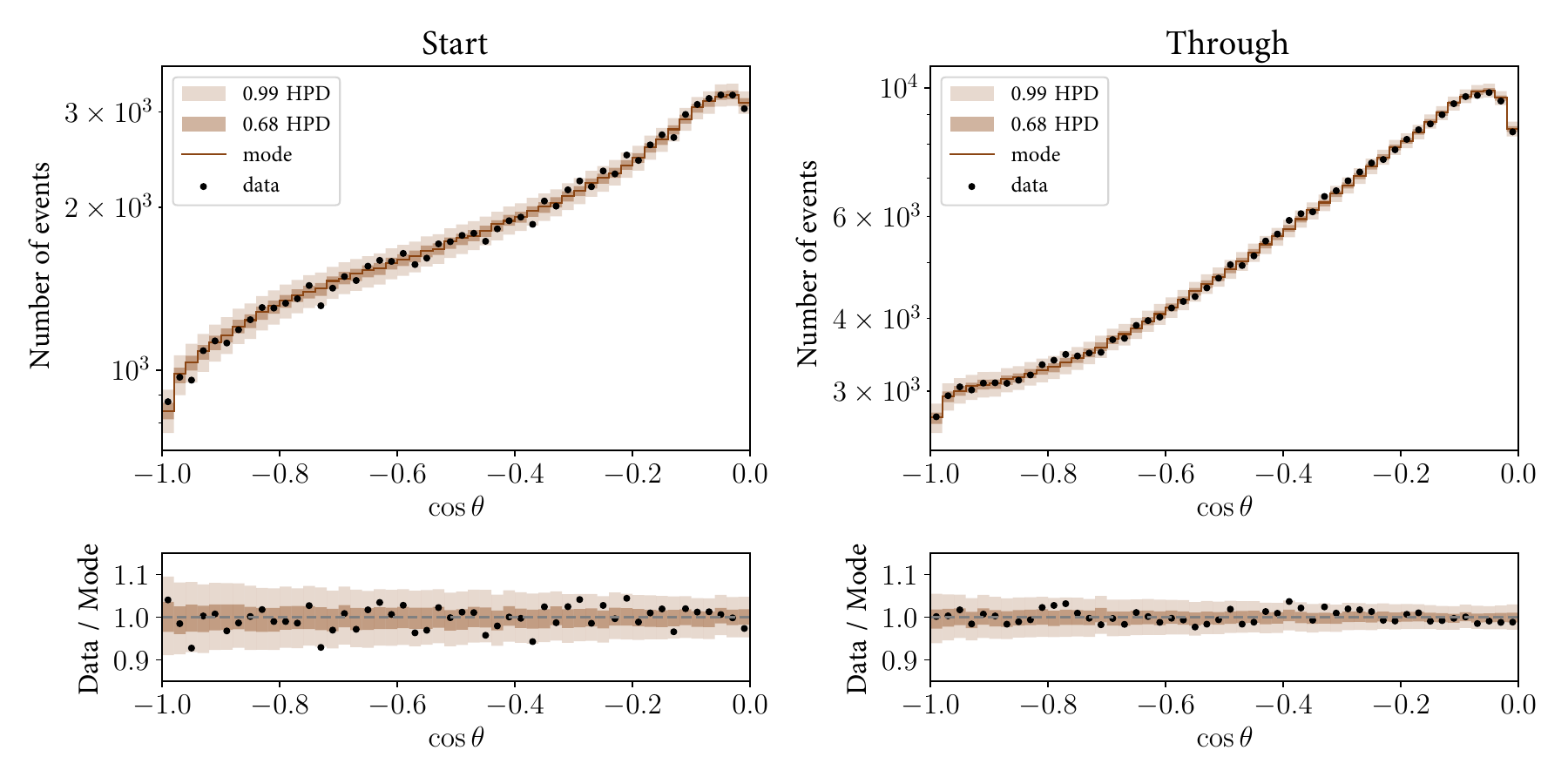}
    \caption{\textbf{The zenith distribution, comparing data and best fit posteriors.} In the top panels, the data (black points) are overlaid with posterior predictive distributions (brown bands showing the mode, 68\% HPD, and 99\% HPD regions). The bottom panels show the ratio relative to the mode of the posterior distribution.}
    \label{fig:zenith_dist}
\end{figure*}

To provide additional goodness-of-fit information beyond the posterior predictive $p$-value reported in the Methods, Figs.~\ref{fig:energy_dist} and \ref{fig:zenith_dist} compare the observed one-dimensional distributions in reconstructed energy and zenith angle to the posterior predictive distributions obtained from the fit. 
In each case, the top panels show the data (black points) overlaid with posterior predictive bands (brown), including the mode and the 68\% and 99\% HPD regions. The bottom panels show ratios relative to the posterior mode, illustrating agreement over the full range and highlighting any localized deviations.

\begin{table}
    \centering
    \footnotesize
    \caption{List of systematic parameters considered in this analysis along with their priors: the center, range, width, and shapes, as well as the posterior mode and 68\% HPD interval. They are split into five main categories.}
\label{tab:systematics}
    \begin{tabular}{lcccccc}
    \toprule
    Parameter (Short name) & Prior center & Prior width ($\pm1\sigma$) & Prior range & Prior shape & Mode & 68\% HPD \\
    \midrule
    \multicolumn{7}{c}{Overall Normalization} \\
    \midrule
    Overall normalization ($\mathrm{Norm}$) & 1.0 & $\pm$0.2 & [0.1, 3] & Gaussian & 1.00 & [0.97, 1.02] \\
    \midrule
    \multicolumn{7}{c}{Detector Parameters} \\
    \midrule
    DOM efficiency ($\mathrm{DOM}_\mathrm{eff}$)  & 1.27 & $\pm$0.1 & [1.234, 1.346] & Gaussian & 1.27 & [1.27, 1.28] \\
    Ice Amplitude 0 ($\mathrm{Ice \:A_0}$) & 0.0 & $\pm$1.0 & [-3, 3] & Gaussian & 0.36 & [-0.13, 0.98] \\
    Ice Amplitude 1 ($\mathrm{Ice \:A_1}$) & 0.0 & $\pm$1.0 & [-3, 3] & Gaussian & 1.38 & [0.84, 1.86] \\
    Ice Amplitude 2 ($\mathrm{Ice \:A_2}$) & 0.0 & $\pm$1.0 & [-3, 3] & Gaussian & 1.34 & [0.84, 1.83] \\
    Ice Amplitude 3 ($\mathrm{Ice \:A_3}$) & 0.0 & $\pm$1.0 & [-3, 3] & Gaussian & 0.62 & [0.12, 1.12] \\
    Ice Amplitude 4 ($\mathrm{Ice \:A_4}$) & 0.0 & $\pm$1.0 & [-3, 3] & Gaussian & 1.01 & [0.41, 1.47] \\
    Ice Phase 1 ($\mathrm{Ice \:Phs_1}$) & 0.0 & $\pm$1.0 & [-3, 3] & Gaussian & $-1.48$ & [$-1.83$, $-1.15$] \\
    Ice Phase 2 ($\mathrm{Ice \:Phs_2}$) & 0.0 & $\pm$1.0 & [-3, 3] & Gaussian & $-0.49$ & [$-0.92$, $-0.03$] \\
    Ice Phase 3 ($\mathrm{Ice \:Phs_3}$) & 0.0 & $\pm$1.0 & [-3, 3] & Gaussian & $-0.25$ & [$-0.72$, 0.34] \\
    Ice Phase 4 ($\mathrm{Ice \:Phs_4}$) & 0.0 & $\pm$1.0 & [-3, 3] & Gaussian & 0.20 & [$-0.25$, 0.73] \\
    Hole Ice & -1.0 & $\pm$10 & [-5.35, 1.85] & Gaussian & 1.74 & [1.02, 1.85] \\
    \midrule
    \multicolumn{7}{c}{Conventional Atmospheric Flux Parameters} \\
    \midrule
    Atmospheric density ($\rho_\mathrm{atm}$) & 0.0 & $\pm$1.0 & [-3, 3] & Gaussian & $-0.98$ & [$-1.73$, $-0.33$] \\
    Kaon energy loss ($\sigma_\mathrm{K-Air}$) & 0.0 & $\pm$1.0 & [-3, 3] & Gaussian & 0.68 & [$-0.19$, 1.56] \\
    K$_{158G}^{+}$ & 0.0 & $\pm$1.0 & [-2, 2] & Gaussian & 1.02 & [0.22, 1.70] \\
    K$_{158G}^{-}$ & 0.0 & $\pm$1.0 & [-2, 2] & Gaussian & 0.20 & [$-0.60$, 1.08] \\
    $\pi_{20T}^{+}$ & 0.0 & $\pm$1.0 & [-2, 2] & Gaussian & 0.20 & [$-0.64$, 0.77] \\
    $\pi_{20T}^{-}$ & 0.0 & $\pm$1.0 & [-2, 2] & Gaussian & 0.12 & [$-0.60$, 0.79] \\
    K$_{2P}^{+}$ & 0.0 & $\pm$1.0 & [-1, 2] & Gaussian & 0.32 & [$-0.01$, 0.62] \\
    K$_{2P}^{-}$ & 0.0 & $\pm$1.0 & [-1.5, 2] & Gaussian & 0.29 & [$-0.06$, 0.59] \\
    $\pi_{2P}^{+}$ & 0.0 & $\pm$1.0 & [-2, 2] & Gaussian & $-1.60$ & [$-2.00$, $-1.10$] \\
    $\pi_{2P}^{-}$ & 0.0 & $\pm$1.0 & [-2, 2] & Gaussian & $-1.18$ & [$-1.84$, $-0.54$] \\
    p$_{2P}$ & 0.0 & $\pm$1.0 & [-2, 2] & Gaussian & $-0.45$ & [$-1.29$, 0.40] \\
    n$_{2P}$ & 0.0 & $\pm$1.0 & [-2, 2] & Gaussian & $-0.29$ & [$-1.07$, 0.62] \\
    GSF$_{1}$ & 0.0 & $\pm$1.0 & [-4, 4] & Gaussian & $-0.26$ & [$-1.12$, 0.70] \\
    GSF$_{2}$ & 0.0 & $\pm$1.0 & [-4, 4] & Gaussian & $-0.11$ & [$-1.13$, 0.72] \\
    GSF$_{3}$ & 0.0 & $\pm$1.0 & [-4, 4] & Gaussian & $-0.07$ & [$-0.95$, 0.90] \\
    GSF$_{4}$ & 0.0 & $\pm$1.0 & [-4, 4] & Gaussian & $-0.11$ & [$-1.11$, 0.74] \\
    GSF$_{5}$ & 0.0 & $\pm$1.0 & [-4, 4] & Gaussian & 1.80 & [1.11, 2.40] \\
    GSF$_{6}$ & 0.0 & $\pm$1.0 & [-4, 4] & Gaussian & $-1.07$ & [$-1.93$, $-0.26$] \\
    \midrule
    \multicolumn{7}{c}{Non-Conventional Flux Parameters} \\
    \midrule
    \shortstack[l]{Astrophysical normalization ($\Phi^\mathrm{HE}$) \\ {[$10^{-18}\,\mathrm{GeV}^{-1}\mathrm{sr}^{-1}\mathrm{s}^{-1}\mathrm{cm}^{-2}$]}} & 0.787 & $\pm$0.36 & [0, 3] & Gaussian & 1.06 & [0.80, 1.31] \\
    $\Delta\gamma_{1}$, tilt from -2.5 ($\Delta\gamma_1^\mathrm{HE}$) & 0.0 & $\pm$0.36 & [-2, 2] & Gaussian & 0.85 & [0.77, 0.95] \\
    $\Delta\gamma_{2}$, tilt from -2.5 ($\Delta\gamma_2^\mathrm{HE}$) & 0.0 & $\pm$0.36 & [-2, 2] & Gaussian & $-0.11$ & [$-0.24$, 0.03] \\
    Log of break energy ($\log_{10}E_\mathrm{break}^\mathrm{HE}/\mathrm{GeV}$) & - & - & [4, 6] & Uniform & 4.27 & [4.11, 4.41] \\
    Prompt normalization ($\mathrm{promptNorm}$) & 0.0 & $\pm$1.0 & [0, 3] & Gaussian & 0.43 & [0.01, 1.02] \\
    \midrule
    \multicolumn{7}{c}{Attenuation cross section parameters} \\
    \midrule
    $\nu$ cross section ($\nu \: \mathrm{Att}$) & 1.0 & $\pm$0.1 & [0.82, 1.18] & Gaussian & 1.01 & [0.94, 1.10] \\
    $\bar{\nu}$ cross section ($\bar\nu \: \mathrm{Att}$) & 1.0 & $\pm$0.1 & [0.82, 1.18] & Gaussian & 1.00 & [0.92, 1.09] \\
    \bottomrule
    \end{tabular}
\end{table}

\subsection{Extensions of main results}

\begin{figure*}
    \centering
    \includegraphics[width=0.7\linewidth]{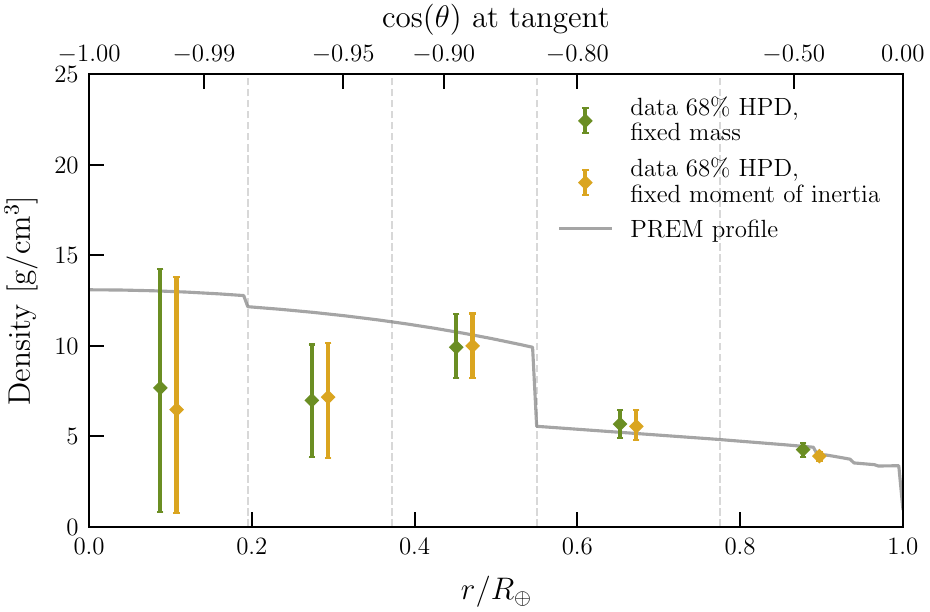}
    \caption{\textbf{Inferred density profile for the five-shell parametrization, equivalent to the measurement shown in Fig.~\ref{fig:5B_profile} but with the mass or moment of inertia as an external constraint on the density parameters.} 
    As before, we show the mode and 68\% HPD ranges as the measurement in the error bars.
    The mass and moment of inertia values used for the constraint are the reference values shown in Fig.~\ref{fig:5B_mass_post} for comparison. 
    The mass is $5.97 \times 10^{24}\,\mathrm{kg}$ and the moment of inertia is $8.01 \times 10^{37}\,\mathrm{kg}\,\mathrm{m}^2$.
    Most notably, under these strong external constraints, the outer layers describing the mantle are significantly more precise compared to the unconstrained measurement in Fig.~\ref{fig:5B_profile}.
    }
    \label{fig:5B_profile_external}
\end{figure*}

Fig.~\ref{fig:5B_profile_external} shows the inferred density profile for the five-shell parametrization, identical in method to that of Fig.~\ref{fig:5B_profile}, but with mass and moment of inertia set as external constraints on the fit.
They are set to the reference values, based on gravitational and seismic measurements, used for comparison in Fig.~\ref{fig:5B_mass_post}. 
In particular, the outer layers show significantly more precision; this is due to the fact that in a spherical body, the mass and moment of inertia are dominated by the density of the outermost layers. 
This result shows that fixing the mass or moment of inertia drives the fit, and completely dominates any neutrino-based contribution to the value of the density measurement in the outer shells.

\begin{figure*}
    \centering
    \includegraphics[width=0.7\linewidth]{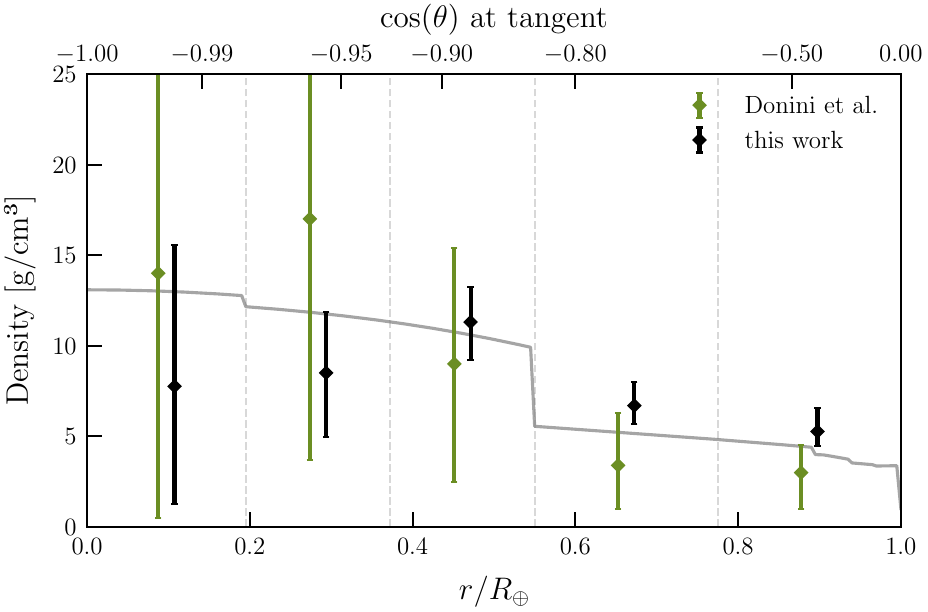}
    \caption{\textbf{Comparison of the five-shell density posteriors from this work and from the previous IceCube neutrino tomography measurement of Ref.~\cite{Donini:2018tsg}.}
    Points indicate posterior modes and error bars give 68\% HPD credible intervals, normalized to the PREM density in each shell.
    The improvement in constraining power across all shells reflects the larger dataset and improved systematic treatment used in this analysis.}
    \label{fig:5B_comparison}
\end{figure*}

Fig.~\ref{fig:5B_comparison} compares the five-shell density posteriors from this work to those reported in Ref.~\cite{Donini:2018tsg}, the previous IceCube-based neutrino tomography measurement.
The improvement in sensitivity is evident across all shells, reflecting the approximately order-of-magnitude increase in event statistics and the significantly updated systematic uncertainty treatment employed here.

\begin{figure*}
    \centering
    \includegraphics[width=0.7\linewidth]{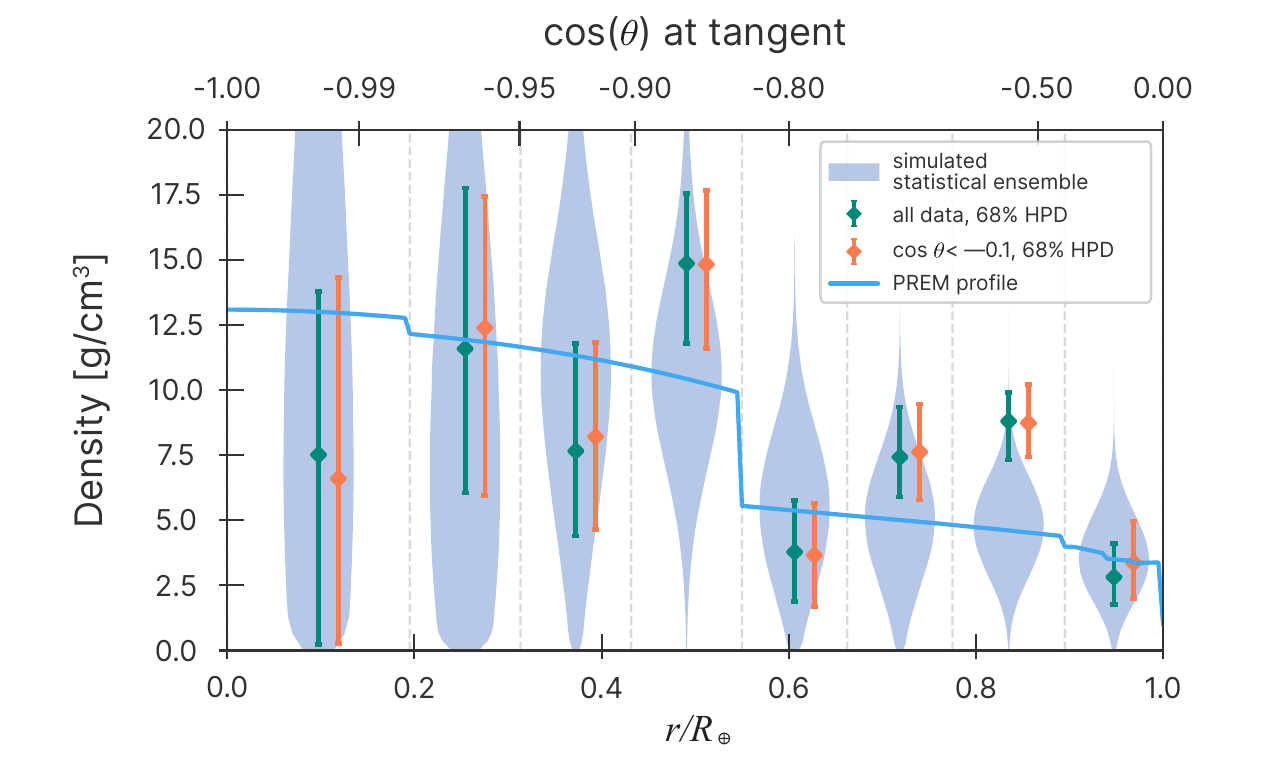}
    \caption{\textbf{Inferred density profile for the eight-shell parametrization, shown as an extension of the five-shell result in Fig.~\ref{fig:5B_profile}.} The format is identical to Fig.~\ref{fig:5B_profile}. 
    The fraction of the marginal posterior lying below the PREM-normalised value of unity for shells 1 through 8 is $65\%$, $50\%$, $80\%$, $7.3\%$, $79\%$, $5.8\%$, $0.050\%$, $71\%$ (all data) and $64\%$, $49\%$, $79\%$, $8.2\%$, $77\%$, $7.6\%$, $0.17\%$, $59\%$ (excl.\ near-horizontal).}
    \label{fig:8B_profile}
\end{figure*}

In Fig.~\ref{fig:8B_profile} we present the inferred density profile for the eight-shell parametrization as a higher-granularity extension of the five-shell result in Fig.~\ref{fig:5B_profile}.
The format is identical to the five-shell presentation, including the comparison to PREM and the two-stage fit with and without the near-horizontal region.
For the all-data fit, we find a posterior mode of $M = 7.17 \times 10^{24}\,\mathrm{kg}$, with a 68\% HPD interval of $[6.22,\, 8.44] \times 10^{24}\,\mathrm{kg}$, and a polar moment of inertia with posterior mode $I = 1.01 \times 10^{38}\,\mathrm{kg}\,\mathrm{m}^2$, with a 68\% HPD interval of $[8.47 \times 10^{37},\, 1.21 \times 10^{38}]\,\mathrm{kg}\,\mathrm{m}^2$, giving a normalized moment of inertia $I/(MR_\oplus^2) \approx 0.346$.

\end{document}